\documentclass[epj]{svjour}
\usepackage{epsfig} \usepackage{psfig}
\begin{document}

\title{Comparative Regge analysis of $\Lambda, \Sigma^0, \Lambda(1520)$ and $\Theta^+$
production in $\gamma p$, $\pi p$ and $pp$ reactions}
\author{
V.Yu.~Grishina $^a$, L.A.~Kondratyuk $^{b}$, W.~Cassing $^c$,
 M. Mirazita $^d$  and P.~Rossi $^d$
\\}
\institute{
$^a$ Institute for Nuclear Research, 60th October
Anniversary Prospect 7A, 117312 Moscow, Russia\\
$^b$ Institute for Theoretical and Experimental Physics, B.\
  Cheremushkinskaya 25, 117218 Moscow, Russia \\
$^c$ Institute for Theoretical Physics,  University of Giessen,
Heinrich-Buff-Ring 16, D-35392 Giessen, Germany\\ $^d$ INFN -
Laboratori Nazionali di Frascati, CP 13, via E. Fermi, 40; I-00044,
Frascati, Italy\\}
\date{}

\abstract{Using the Quark-Gluon Strings Model -- combined with Regge
phenomenology -- we perform a comparative analysis of  $\Lambda$,
$\Sigma^0$, $\Lambda(1520)$ and $\Theta^+$ production in binary
reactions induced by photon, pion and proton beams on the nucleon.
We find that the existing experimental data on the $\gamma p \to K^+
\Lambda$ differential and total cross sections can be described very
well by the model for photon energies $1 - 16$~GeV and $-t <
2$~GeV$^2$ assuming a dominant contribution of the $K^*$ Regge
trajectory. Moreover, using the same parameters we also reproduce
the total $\gamma p \to K^+ \Sigma^0$ and $\gamma p \to K^+
\Lambda(1520)$ cross sections suggesting a 'universality' of the
Regge model. In order to check the consistency of the approach we
evaluate the differential and total cross sections for the
reaction~$\pi^- p \to K^0 \Lambda$ which is also found to be
dominated by the $K^*$ Regge trajectory. Using the apparent
'universality' of the Regge model we extend our scheme to the
analysis of the binary reactions $\gamma p \to \bar{K}^0 \Theta^+$,
$\pi^- p \to K^- \Theta^+$ and $pp \to \Sigma^+ \Theta^+$ as well as
the exclusive and inclusive $\Theta^+$ production in the reactions
$pp \to p \bar{K}^0 \Theta^+$ and $pp \to \Theta^+ X$. Our detailed
studies demonstrate that $\Theta^+$ production does not follow the
'universality' principle thus suggesting an essentially different
internal structure of the exotic baryon relative to conventional
hyperons or hyperon resonances.}

\PACS{13.75.Gx,13.75.Jz,12.39.Mk}
\authorrunning{V.Yu. Grishina et al.}

\titlerunning{Comparative Regge analysis ...}
\maketitle

\newpage

\section{Introduction}
Inspite of the belief that the structure of baryons in the octet and
decuplet representation is roughly understood (and exhausted),
recent claims on the discovery of the manifestly exotic baryon
$\Theta^+$  have opened a new chapter in hadron physics (see e.g.
Refs.~\cite{Nakano,Barmin,Stepanyan,Barth,HERMES,Kubarovsky,Asratyan,Alt,Kubarovsky2,Aleev,Abdel-Bary}).
Although the existence and properties of this exotic and long-lived
baryonic state still need final experimental confirmation a variety
of theoretical models have been set up with different assumptions
about the internal structure of the $\Theta^+$. Here the
quark-soliton model of Diakonov, Petrov and Polyakov \cite{Diakonov}
was the first to claim the existence of an antidecuplet with rather
narrow spectral width (actually prior to the experimental
observations). However, the $\Theta^+$ might also be 'bound' due to
strong diquark correlations (in a relative $p$-wave) as proposed in
Ref. \cite{Jaffe} or due to a strong mixing with the
octet~\cite{Hyodo}. Alternatively, it might even be explained on the
basis of the constituent quark model involving clusters \cite{Oka}.
Further models have been proposed in the last 2 years that all claim
a different dynamical origin of the pentaquark $\Theta^+$
\cite{Beane,Ioffe} (cf. \cite{Jafferev} and Refs. cited therein).
Additionally, the properties of the exotic state have been analysed
within the framework of QCD sum rules \cite{Sugiyama} and even
lattice QCD \cite{Fodor,Bali,LQCDlast,Csikor}.

However, for a better understanding of this exotic state and its
wave function (in terms of the elementary degrees of freedom)
it is very important to study the dynamics
of $\Theta^+$ production in comparison to the
production of non-exotic strange baryons. In this respect
exclusive reactions with strangeness ($s\bar{s}$) production are of interest, i.e.
(starting with $\gamma$ induced reactions):
\begin{equation}
\gamma p \to \bar{K}^0 \Theta ^+ \ , \label{gpKTheta}
\end{equation}
\begin{equation}
\gamma p \to \bar{K}^{*0} \Theta ^+ \label{gpK*Theta}
\end{equation}
and
\begin{equation}
\gamma d \to \Lambda \Theta ^+ \ .
\label{gdLTheta}
\end{equation}
The first two reactions (\ref{gpKTheta}) and (\ref{gpK*Theta}) -- where
the $s$ quark ends up in the mesonic final state -- can be compared with
 $\Lambda$ production in the binary reactions
\begin{equation}
\gamma p \to  K^+ \Lambda,
\label{gpKL}
\end{equation}
\begin{equation}
\gamma p \to  K^{*+}  \Lambda
\label{gpK*L}
\end{equation}
where the $s$ quark ends up in the hyperon.
Note that  very detailed measurements of the
reaction (\ref{gpKL}) have been performed in the energy range from
threshold up to a photon energy of 2.6 GeV \cite{Glander04}. The
third reaction (\ref{gdLTheta}), furthermore, can be compared with the two-body
deuteron photodisintegration reaction $\gamma d \to p n $ which
has been studied recently at Jlab \cite{Rossi}.

We recall that several studies of $\Theta^+$ photoproduction have
been performed within the framework of isobar models using the Born
approximation \cite{Oh,Nam,Li,Ko,Liu2,Yu}. Since these models
involve a variety of uncertain parameters (coupling constants and
cutoff's)  the resulting cross sections differ from several nb to
almost 1~$\mu$b. On the other hand the  Regge model has a
substantial advantage  that the amount of uncertain parameters is
much lower and that the latter can be fixed by other reactions in a
more reliable fashion \cite{Mart}. Accordingly, in this work  we
will apply the Quark-Gluon Strings Model (QGSM) combined with Regge
phenomenology to the analysis of the differential and total cross
sections of the exclusive reactions $\gamma p \to \ K^+ \Lambda$,
$\gamma p \to \ K^+ \Sigma^0$
 and $\gamma p \to \bar{K}^0 \Theta^+$. Similar final channels
 will be investigated also for pion and proton induced reactions.

We note that the QGSM was originally proposed by Kaidalov in Ref.
\cite{Kaidalov} for the description of binary hadronic reactions; as
demonstrated in Refs. \cite{Kaidalov,KaidalovSurveys} the QGSM
describes rather well the experimental data on exclusive and
inclusive hadronic reactions at high energy. More recently this
model has been also successfully applied to the description of the
nucleon and pion electromagnetic form factors \cite{Tchekin} as well
as  deuteron photodisintegration  \cite{Rossi,Grishina,Grishina2}.

\begin{figure}
 \begin{center}
    \leavevmode
    \psfig{file=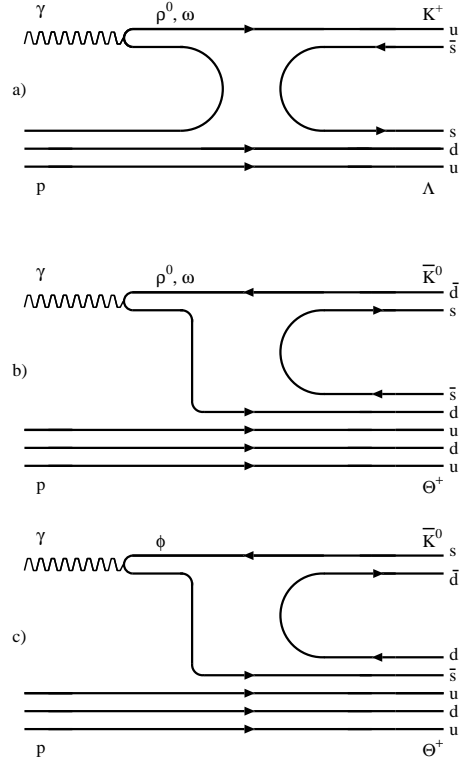,width=6.cm}
    \caption{Quark planar diagrams describing the binary reactions:
 $\gamma p \to \ K^+ \Lambda$ (a), $\gamma p \to \bar{K}^0 \Theta^+$
(b and c).}
\label{fig:qdiagr}
\end{center}
\end{figure}

We recall that the QGSM is based on two ingredients: i) a
topological expansion in QCD and ii) the space-time picture of the
interactions between hadrons, that takes into account the
confinement of quarks.  The $1/N$ expansion in QCD (where $N$ is
the number of colors $N_c$ or flavors $N_f$) was proposed by 't
Hooft \cite{Hooft}; the behavior of different quark-gluon graphs
according to their topology, furthermore, was analyzed by
Veneziano \cite{Veneziano} with the result that in the large $N$
limit the planar quark-gluon graphs become dominant. This approach
-- based on the $1/N_f$ expansion \cite{Veneziano} with $N_c \sim
N_f$  -- was used by Kaidalov \cite{Kaidalov,KaidalovSurveys} in
formulating the QGSM. Again for sufficiently large $N_f$ the
simplest planar quark-gluon graphs were found to give the dominant contribution
to the amplitudes of binary hadronic reactions. Moreover, it can be
shown that (in the space-time representation) the dynamics described
by planar graphs corresponds to the formation and break-up of a
quark-gluon string (or color tube) in the $s$-channel (see e.g.
\cite{Casher,Artru,Casher2,Andersson,Gurvich}). On the other hand an
exchange of the $u$ and $\bar s$ quarks in the $t$-channel
implies that the energy behavior of the amplitudes -- described by
quark diagrams in Fig.~\ref{fig:qdiagr} -- is given by the
contribution of the $K^*$ Regge trajectory. In this sense the QGSM
can be considered as a microscopic model of Regge phenomenology.
This in turn allows to obtain many relations
between amplitudes of different binary reactions and residues of
Regge poles which determine these amplitudes
\cite{Kaidalov,KaidalovSurveys,Boreskov}.

Our investigation is organized as follows: In Section 2 we outline
our approach and present the results for the differential cross
sections $\gamma p \to K^+ \Lambda,\ K^+ \Sigma^0,\ \bar{K}^{0}
\Theta^+$. In Section 3 we compare total cross sections for the
reactions $\gamma p \to K^+ \Lambda(1520)$ and $\gamma p \to
\bar{K^0} \Theta ^+$ while in Section 4 we step on with the pion
induced reactions $\pi^- p \to K^0 \Lambda$ and $\pi^- p \to K^-
\Theta ^+$. An analysis of $\Theta^+$ production in exclusive and
inclusive $NN$  collisions is presented in Section 5 while a summary
of our studies is given in Section 6.

\section{The reactions $\gamma p \to \ K^+ \Lambda$,
$\gamma p \to \ K^+ \Sigma^0$ and $\gamma p \to \bar{K}^{0}
\Theta^+$} 
\begin{figure}
 \begin{center}
    \leavevmode
    \psfig{file=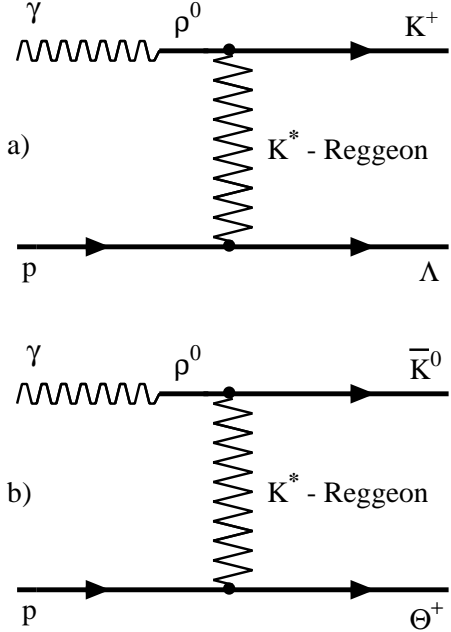,width=6.cm}
    \caption{$K^*$ Reggeon exchanges corresponding to the
quark planar diagrams of
 Fig. \ref{fig:qdiagr}.} \label{fig:Regdiagr}
  \end{center}
\end{figure}

We first concentrate on $\gamma$ induced reactions and
work out the Regge model in more detail. The reaction $\gamma p \to
\ K^+ \Lambda$ can be described by the exchange of two valence ($u$
and $\bar s$) quarks in the $t$-channel with any number of gluon
exchanges between them  (Fig.~\ref{fig:qdiagr} a) ). Alternatively,
in terms of the Regge phenomenology this diagram corresponds to the
$K^*$- Reggeon exchange mechanism shown in
Fig.~\ref{fig:Regdiagr}~a).

Employing the Regge model further on we can write the $\gamma p \to
\ K^+ \Lambda$ amplitude in the form
\begin{eqnarray}
&& T(\gamma p \to  K^+ \Lambda)
\simeq \frac{e}{2 \gamma_{\rho}} T(\rho^0 p \to  K^+ \Lambda) =
\nonumber \\
&& \frac{e}{2 \gamma_{\rho}}\ g_{\rho K K^*} \
g_{pK^* \Lambda} \ F_1(t) \ (-s/s_0^{K \Lambda})^{\alpha_{K^*}(t)} \ .
\label{gpKLf}
\end{eqnarray}
Here $e^2/4\pi$ is the fine structure constant, $\gamma_{\rho}^2 /4
\pi$ = 0.55, ${\alpha_{K^*}(t)}$ is the $K^*$ Regge trajectory,
$s_0^{K\Lambda}= (M_{\Lambda} + m_{K})^2$, $g_{\rho K K^*}$ and
$g_{pK^* \Lambda}$ are the coupling constants describing the
interaction of the $K^*$ Reggeon with the $ \rho K$ and $p \Lambda$
systems. Within the Reggeized-Born-term model (see eg.
Refs.~\cite{Irving,Levy,Guidal}) it is assumed that the coupling
constants $g_i$ in the Regge amplitude of Eq.~(\ref{gpKLf}) can be
identified with the coupling constants in an effective Lagrangian
model. However it is difficult to justify this assumption and we do
not address this model here. We follow another approach for a
'Reggeization of the amplitude' as proposed in Refs.
\cite{Tchekin,Grishina,Grishina2}, i.e. by using the $s-$channel
convolution representation in the QGSM. In this approach one can
express the amplitude for the reaction $\gamma p \to \ K^+ \Lambda$
in terms of the $s$-channel convolution of two amplitudes: $
T(\gamma p \to q + qq)$ and $T(q+qq  \to  K^+ \Lambda)$ (see
Fig.~1). Then -- using the Regge representation for the hadron-quark
and quark-hadron transition amplitudes -- we can Reggeize the binary
amplitude  $\gamma p \to \ K^+ \Lambda$. Such a procedure was
applied in Refs. \cite{Grishina,Grishina2} to define the spin
structure of the deuteron photoproduction amplitude. We point out
that this approach is more general and gives a vertex structure of
the amplitude at negative $t$ different from the Reggeized-Born
term.

Assuming, furthermore,  that the $\Theta^+$ is a pentaquark of
structure ($uudd\bar{s}$) we can use a similar strategy for the
$\gamma p \to \bar{K}^0 \Theta^+$ reaction. The relevant quark
diagrams for this reaction are shown in  Fig. \ref{fig:qdiagr}  b),
c). It is obvious that in terms of the Regge phenomenology we can
also use the $K^*$- Reggeon exchange model to describe the reaction
$\gamma p \to \bar{K}^0 \Theta^+$ (cf. Fig.~\ref{fig:Regdiagr}~b).
The $\gamma p \to \bar{K}^0 \Theta^+$ amplitude reads accordingly
\begin{eqnarray}
&& T(\gamma p \to  \bar{K}^0 \Theta^+ ) \simeq \frac{e}{2
\gamma_{\rho}} T(\rho_0 p \to  \bar{K}^0 \Theta^+ )
= \nonumber \\
&&\frac{e}{2 \gamma_{\rho}} \ g_{\rho K K^*}
g_{pK^* \Theta} \ F_2(t) \  (-s/s_0^{K \Theta})^{\alpha_{K^*}(t)} \
\label{gpKThetaf}
\end{eqnarray}
with $s_0^{K\Theta}= (M_{\Theta} + m_{K})^2$. In the following
calculations the form factor squared $|F_i|^ 2 $ in (\ref{gpKLf}),
(\ref{gpKThetaf}) is chosen always in the form
\begin{equation}
|F_i|^2 = (1 - B_i t) \, \exp (2 R_i^2 t).
\label{FF}
\end{equation}

For the further developments it is important to recall that the QGSM
originally was formulated for small scattering angles (or small
negative 4-momentum transfer (squared) $-t$). Thus the question
arises about the extrapolation of the QGSM amplitudes to large
angles (or large $-t$). Here we adopt the same concept as in our
previous works \cite{Grishina,Grishina2}: following Coon et
al.~\cite{Coon} we assume that only a single analytic Regge term
with a logarithmic trajectory gives the dominant contribution to
large momentum transfer processes. As  shown in \cite{Coon} such a
model (denoted as 'logarithmic dual model') can describe very well
the differential cross section $d\sigma/dt$ for elastic $pp$
scattering in the energy range of $5-24$ GeV/c for $-t$ up to
18~GeV$^2$.
The logarithmic Regge trajectory itself can be written in the form
\begin{equation} \label{nonlin}
\alpha(t)= \alpha(0)- (\gamma \nu) \ln (1 - {t}/{T_B}).
\end{equation}
with the parameters $\alpha (0)=0.32$ and $T_B = 6$~GeV$^2$ that
have been fixed in  Refs.\cite{Volkovitsky94,Burak}. To describe the
energy dependence of the $\gamma p \rightarrow K^+ \Lambda$
differential cross section at fixed $t$ we have found $\gamma \nu
=2.75$.

We note in passing that logarithmic Regge trajectories have also been
discussed in Refs. \cite{Bugrij,Ito,Chikovani}. The special limit
$\gamma \nu \to 0$ at large $-t$ corresponds to
'saturated' trajectories, i.e. all trajectories approach a
constant asymptotically. Such a case leads to the 'constituent-interchange
model' that can be considered as a predecessor of the 'asymptotic quark
counting rules' \cite{Brodsky,Hiller}. Moreover, the model with 'saturated'
trajectories has also successfully been applied to
the large-$t$ behavior of exclusive photon- and hadron-induced
reactions in Refs. \cite{Guidal,Fiore,White,Battaglieri}.

Formally, the amplitude (\ref{gpKLf}) does not contain spin variables.
Nevertheless it can be used for a description of the differential
cross section that is averaged over the spin states
of the initial particles and summed up over the polarizations
of the final particles,
\begin{eqnarray}
&&\displaystyle \frac{d\sigma_{\gamma p \to K^+ \Lambda  }}{d t} =
\frac{1}{64\,\pi s}\ \frac{1}{(p_{\pi}^{\mathrm{cm}})^2}\
 \times \nonumber \\
&&\frac{1}{4} \sum_{\lambda_{\gamma}, \lambda_{p}, \lambda_{\Lambda}}
\ \left| \langle \lambda_{\Lambda} | T_{\gamma p \to K^+ \Lambda}(s,t) |
\lambda_{p}, \lambda_{\gamma} \rangle \right|^2 \ .
\end{eqnarray}
Here the amplitude squared can be written as
\begin{eqnarray}
&&\displaystyle
\frac{1}{4} \sum_{\lambda_{\gamma}, \lambda_{p}, \lambda_{\Lambda}}
\ \left| \langle \lambda_{\Lambda} | T_{\gamma p \to K^+ \Lambda}(s,t) |
\lambda_{p}, \lambda_{\gamma} \rangle \right|^2 = \nonumber \\
&& \frac{e^2}{4 \gamma_{\rho}^2}g_{\rho K K^*}^2 \:
g_{pK^* \Lambda}^2 |F_1(t)|^2
\left|-s/s_0^{K \Lambda}\right|^{2 \alpha_{K^*}(t)} \ .
\end{eqnarray}
Let us now discuss constraints that have to be fulfilled for
the residues and coupling constants. In line with
Refs.~\cite{KaidalovSurveys,Volkovitsky94,Nogteva} we assume
that for the planar quark diagrams with light quarks there is
some kind of 'universality' of the secondary Reggeon couplings to
$q \bar q$ mesons involved in a  binary reaction, i.e. in particular
\begin{equation}
g_{\rho K K^{*}}\simeq  g_{\pi K K^*}  \simeq g_{\rho \pi \pi} =
g_0
\label{univers}
\end{equation}
with $g_0 \simeq$ 5.8. Taking $g_{\rho K K^*}=5.8$ and
normalising the differential cross section of the reaction
$\gamma p \to K^+ \Lambda$ at $t=0$ we find
$g_{p K^* \Lambda} \simeq$ 3.5. This result shows that the Reggeon couplings
to mesons and baryons might be, in general, different by up to a factor of 2.

We mention  that the form factor $F_i$ -- determining the $t$-dependence
of the residue -- was parametrized in Refs.~\cite{KaidalovSurveys,Volkovitsky94}
as
\begin{equation} \label{dual}
F_i(t)= \Gamma(1-\alpha_i(t)).
\end{equation}
Indeed, such a choice of the form factor is convenient for an analytical
continuation of the amplitude to positive $t$ where
the $\Gamma$ function decreases exponentially with $t$.
However, in the region of negative $t$ the parametrization
(\ref{dual}) exhibits a factorial growth and is not acceptable (see e.g. the
discussion in Ref. \cite{Cassing}). Accordingly we use the
parametrization of the form factor (\ref{FF}), which decreases
with $t$. To keep the same normalization of the amplitude at $t=0$
we have to change the coupling constant squared as
\begin{equation} \label{g_M}
g_0^2 \to g_M^2=g_0^2 \Gamma(1-\alpha_i(0)),
\end{equation}
where for the $K^*$ trajectory we have $\Gamma(1-\alpha_{K^*}(0)) \simeq
1.32$.

\begin{figure}[t]
  \centerline{\psfig{file=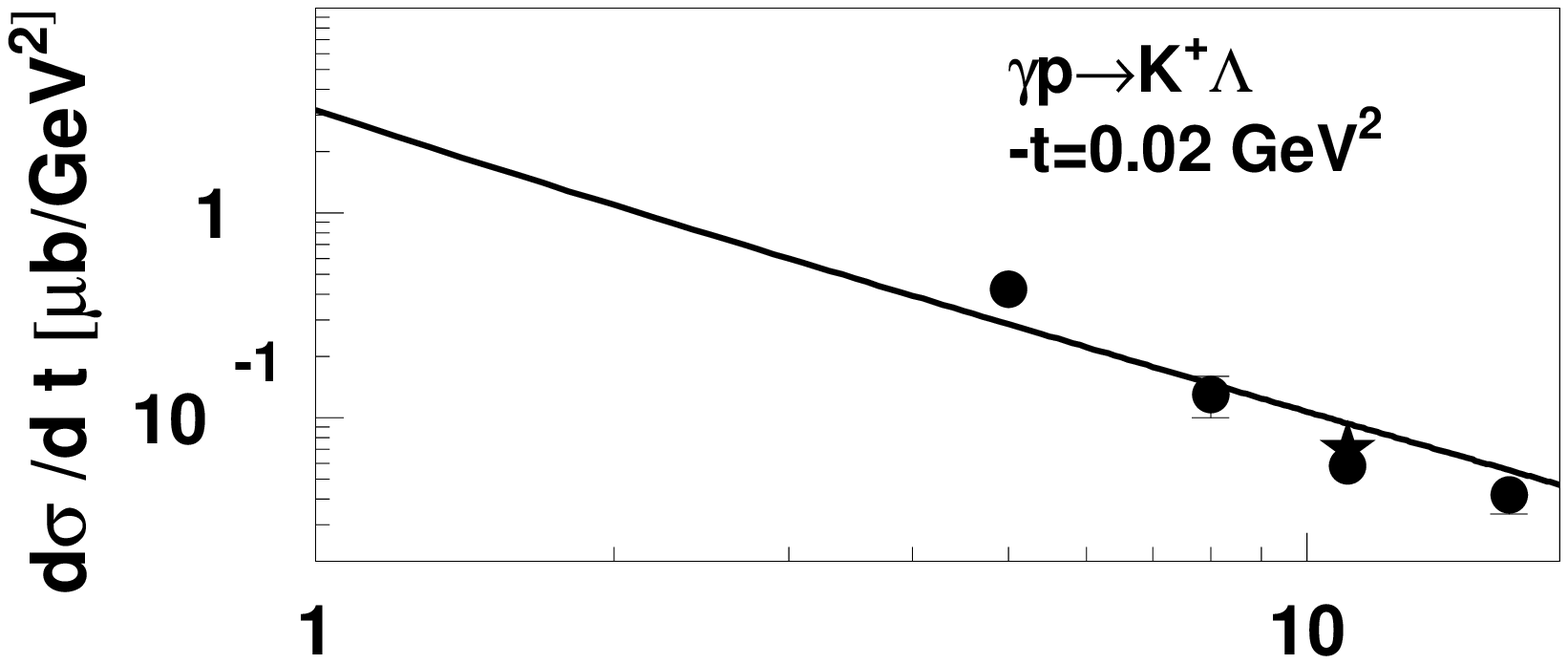,width=7.0cm}}
  \centerline{\psfig{file=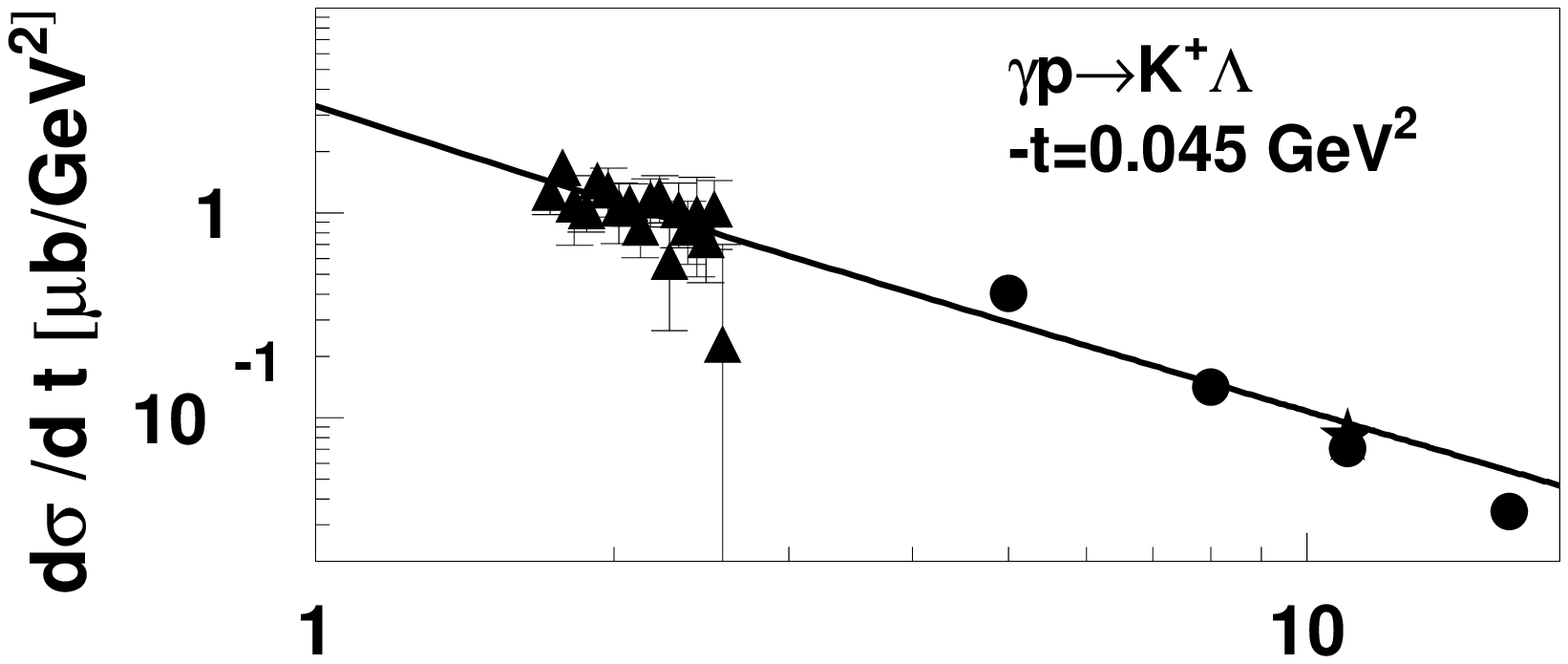,width=7.0cm}}
  \centerline{\psfig{file=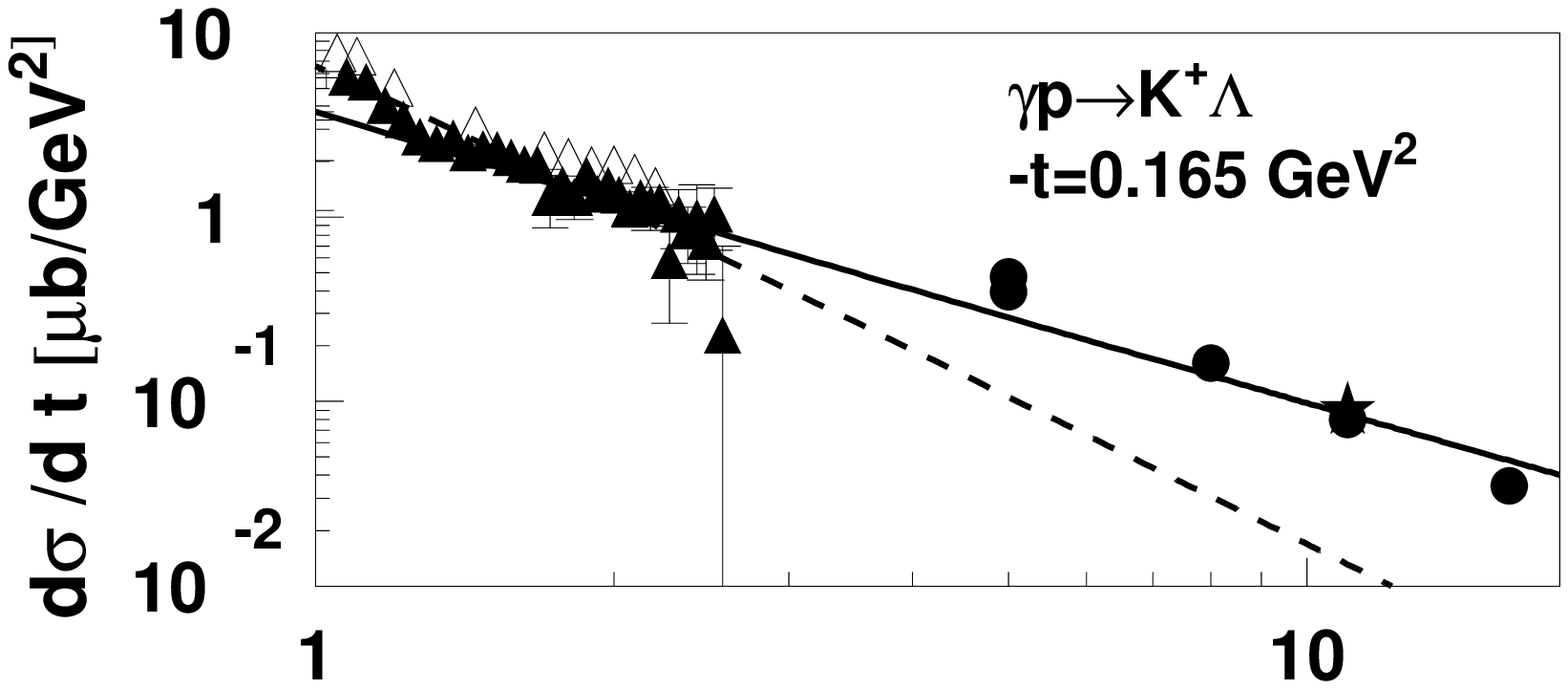,width=7.0cm}}
  \centerline{\psfig{file=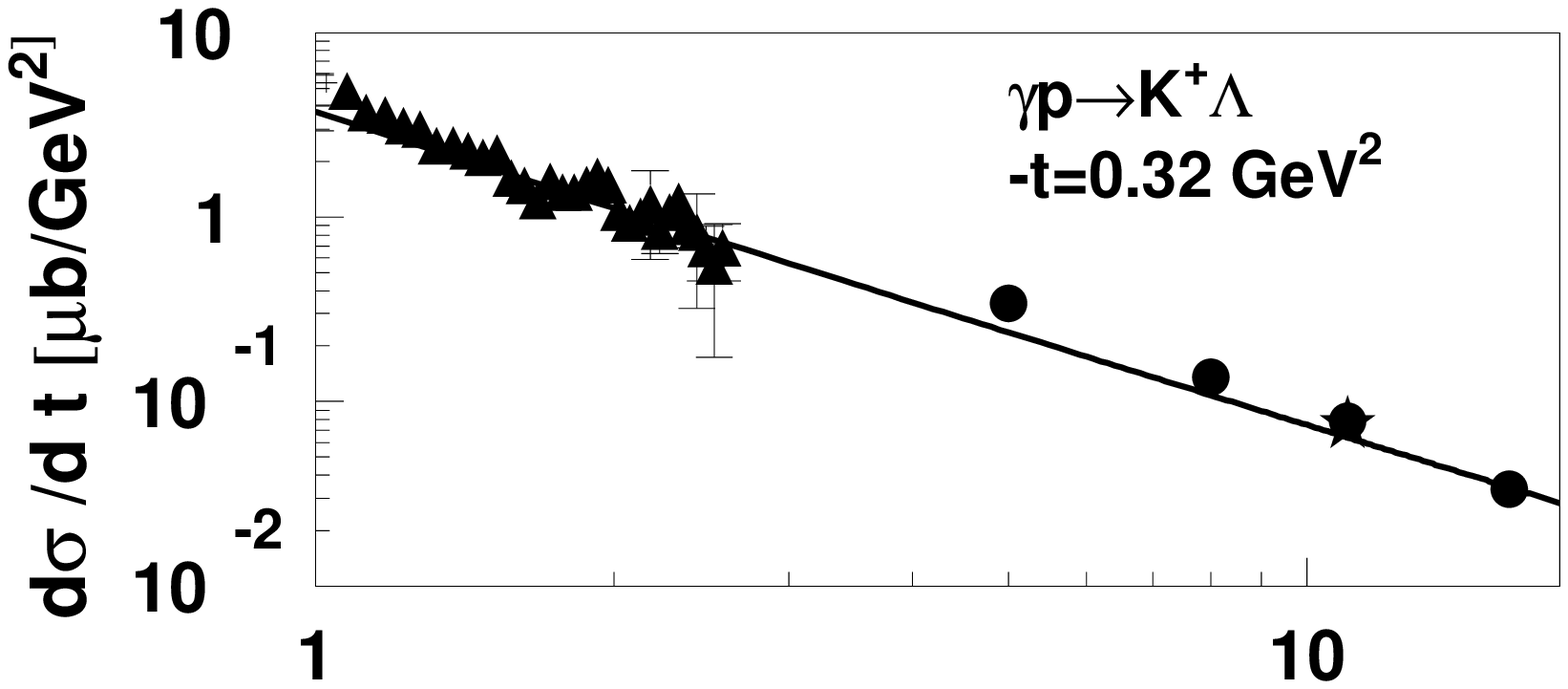,width=7.0cm}}
  \centerline{\psfig{file=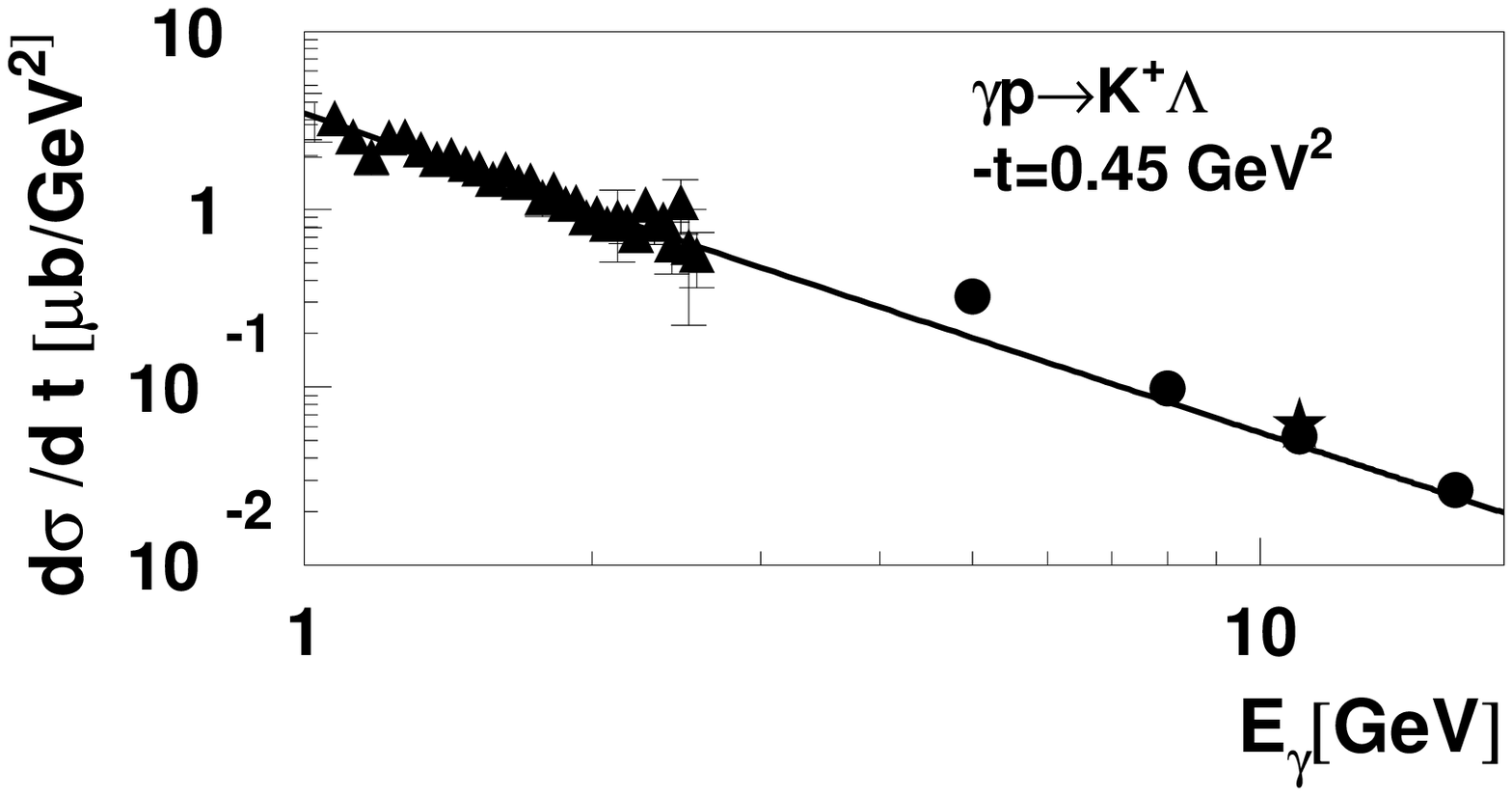,width=7.0cm}}
    \caption{Differential cross section of the reaction
$\gamma p \to  K^+ \Lambda$ at $t=-0.02,-0.045,-0.165,-0.32$ and
--0.45 GeV$^2$ as a function of the laboratory photon energy. The
experimental data are from Refs.\cite{Glander04} (full triangles),
\cite{Boyarski69} (full circles),\cite{Boyarski71}(full stars), and
\cite{Feller72}(empty triangles). The solid line is the result of
the QGSM for the contribution of the $K^*$ logarithmic Regge
trajectory defined by Eq. (\ref{nonlin}). The dashed line (for $t =
-0.165$ GeV$^2$) describes the result for the $K$ Regge trajectory
$\alpha_K(t)=0.7 (t - m_K^2)$ normalized to the data of Ref.
\cite{Glander04} at $E_{\gamma}$ = 2 GeV.  }
  \label{dstklam045}
\end{figure}

\begin{figure}[t]
  \centerline{\psfig{file=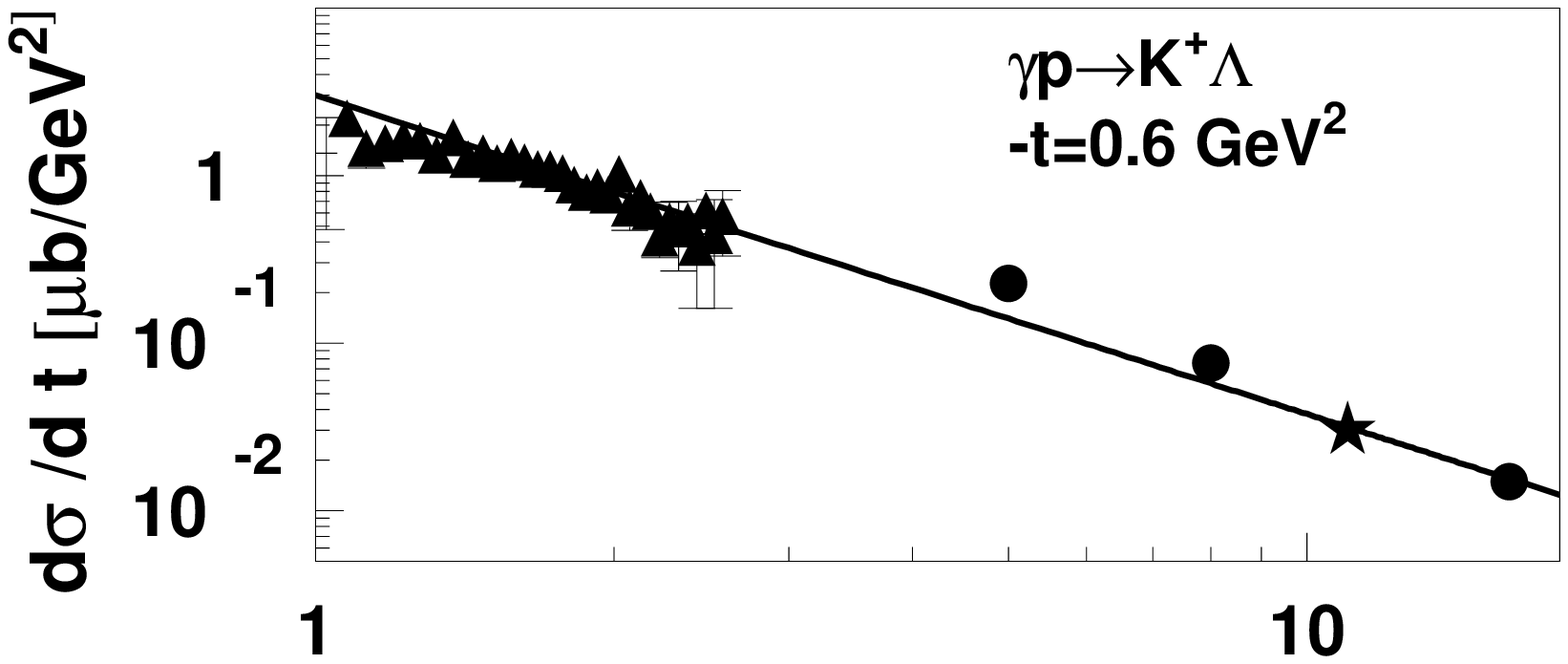,width=7.0cm}}
  \centerline{\psfig{file=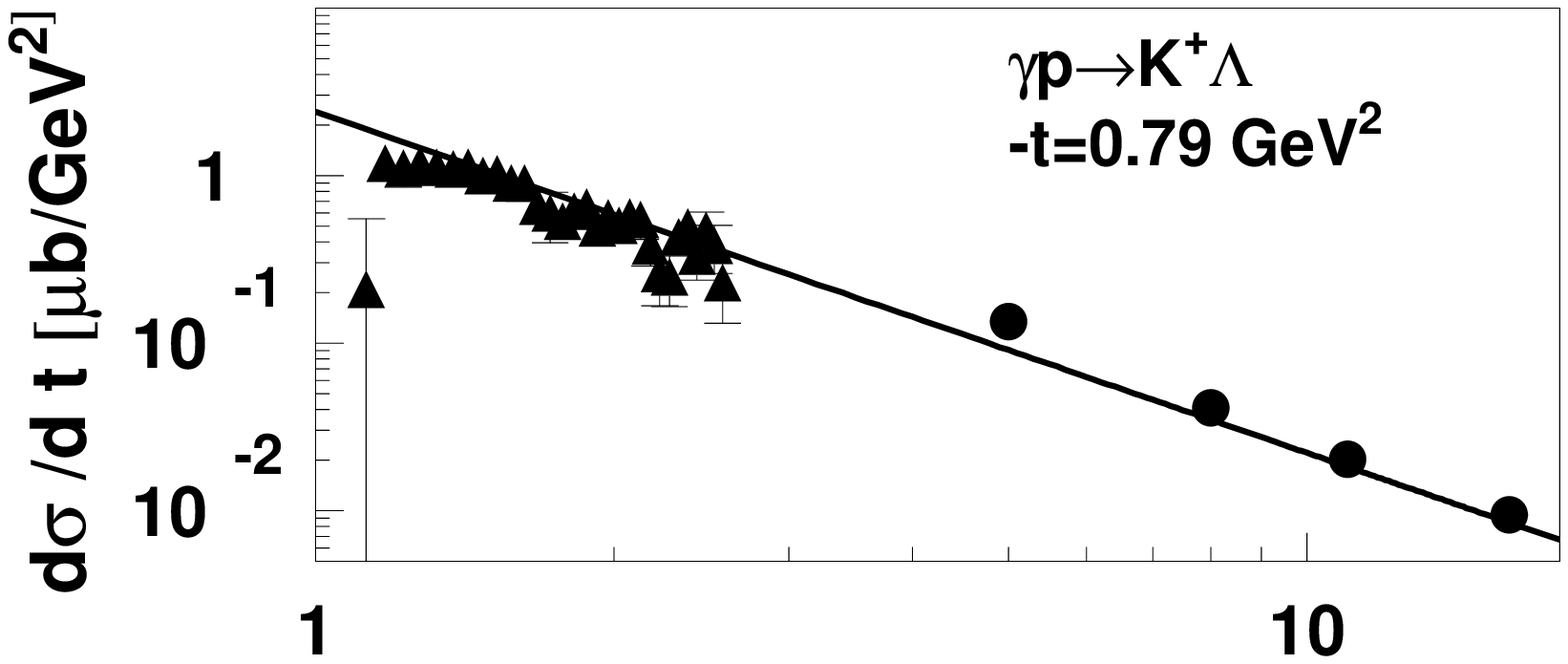,width=7.0cm}}
  \centerline{\psfig{file=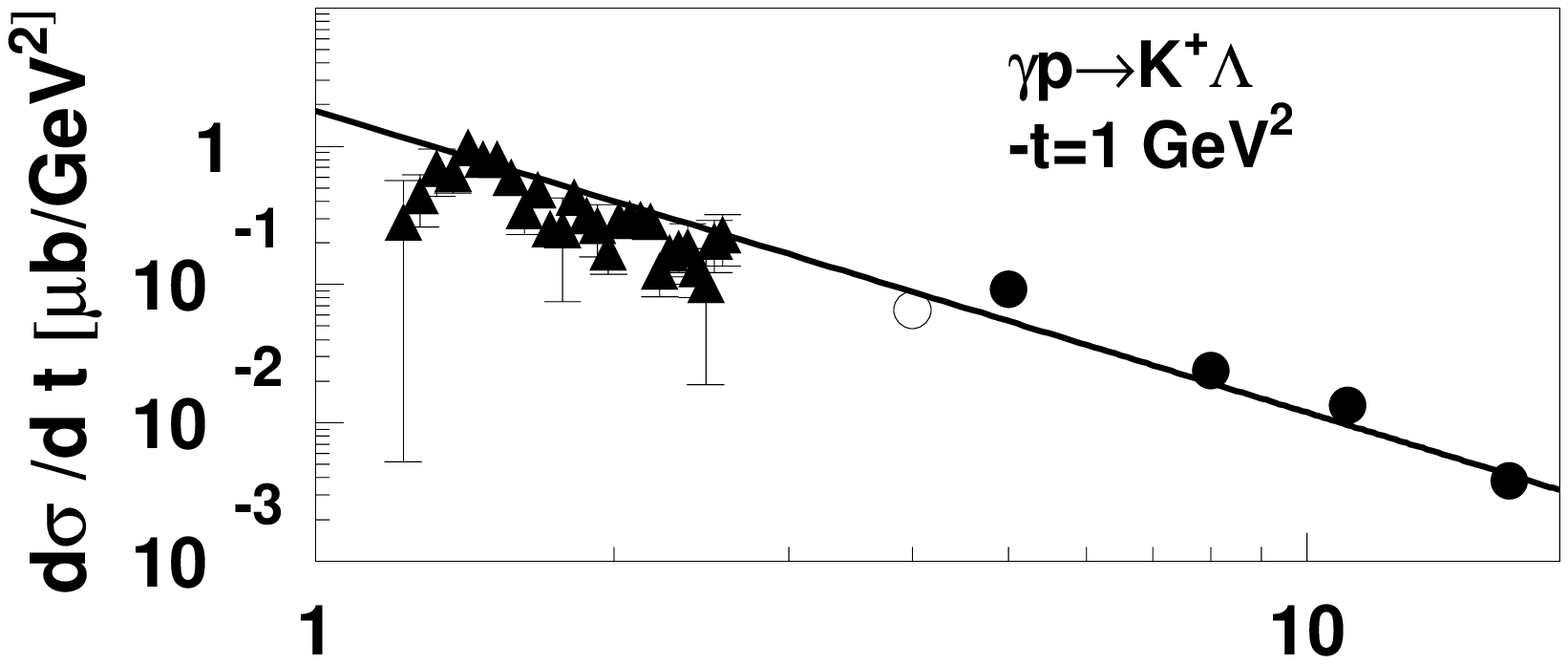,width=7.0cm}}
  \centerline{\psfig{file=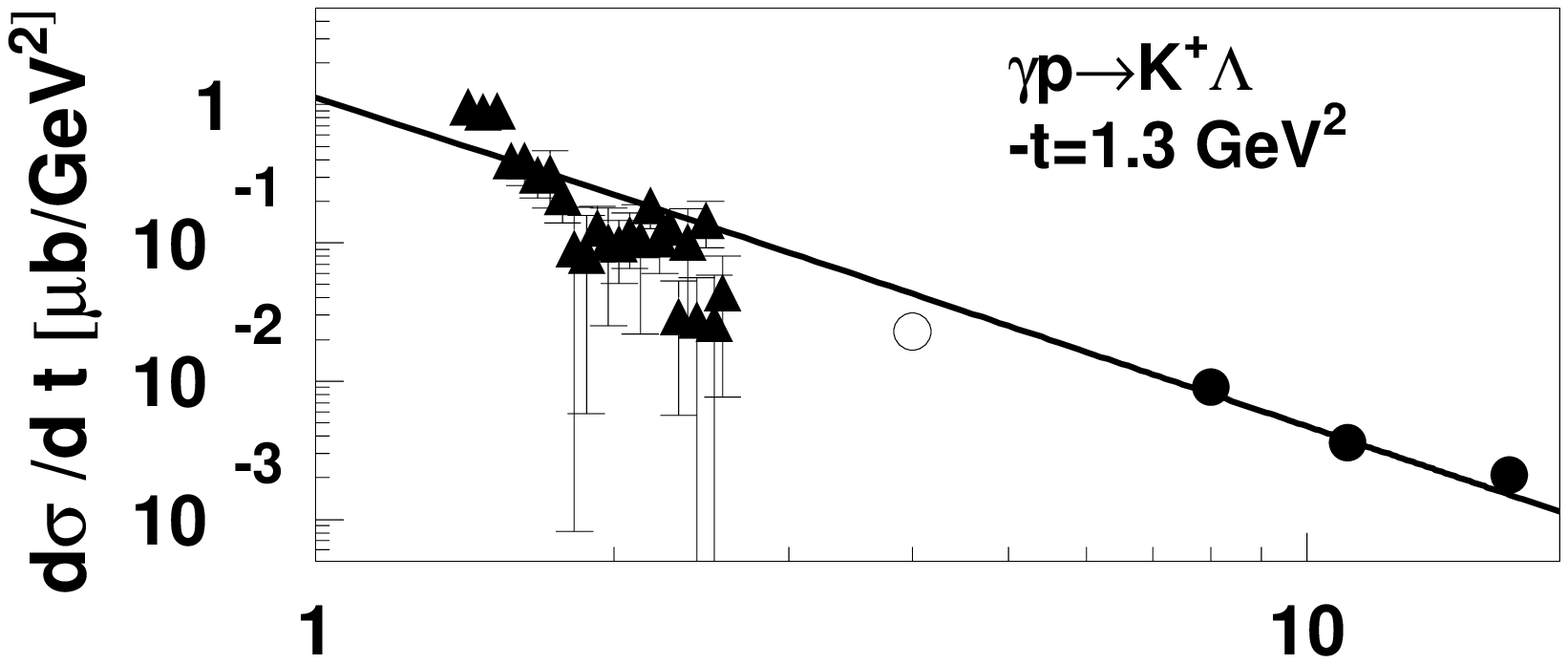,width=7.0cm}}
  \centerline{\psfig{file=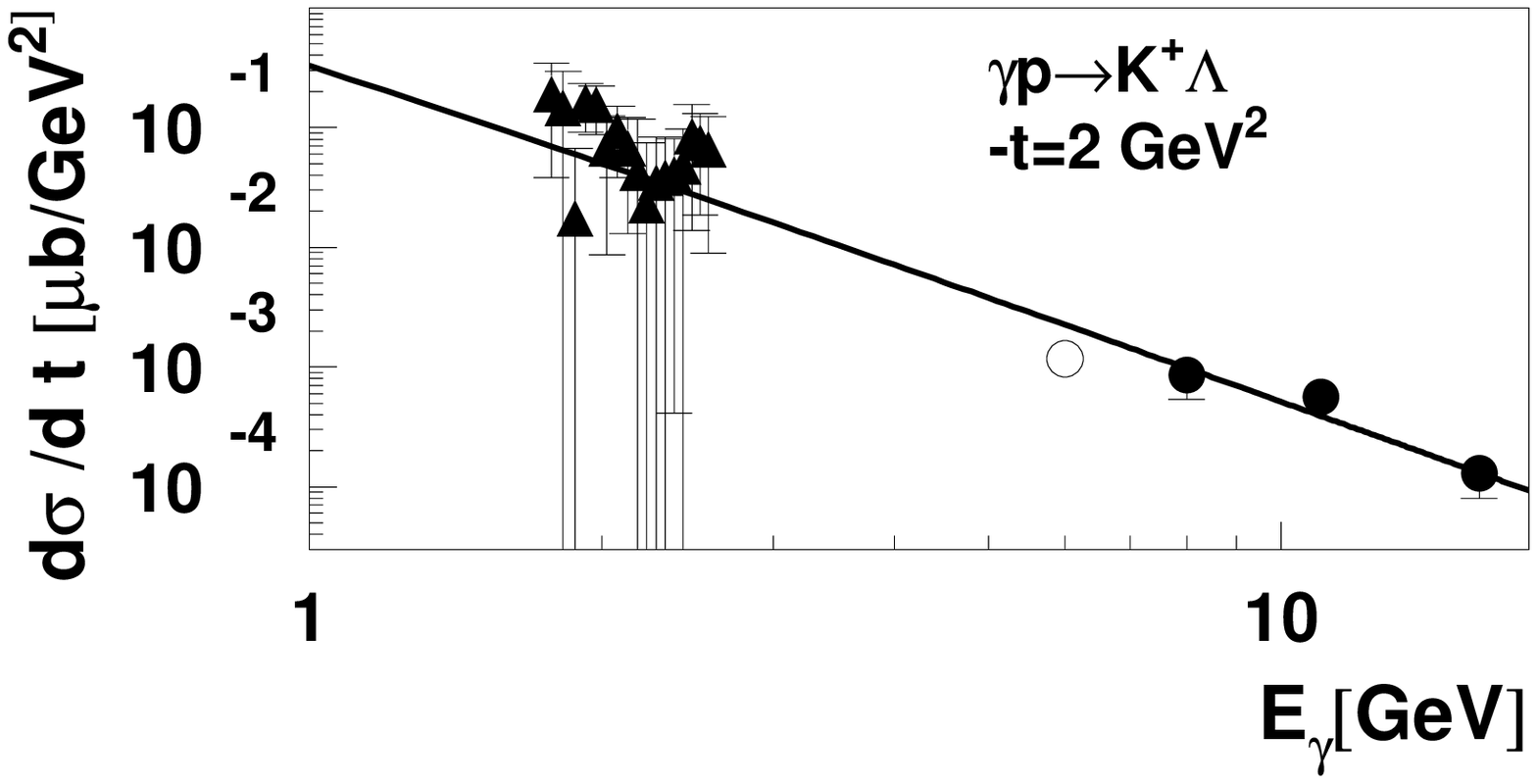,width=7.0cm}}
    \caption{Differential cross section of the reaction
$\gamma p \to  K^+ \Lambda$ at $t=-0.6,-0.79,-1.0,-1.3$ and --2.0
GeV$^2$ as a function of the laboratory photon energy. The
experimental data are from Refs. \cite{Glander04} (full triangles),
\cite{Boyarski69} (full circles), \cite{Boyarski71}(full stars), and
 \cite{Anderson76}(empty circle).
The solid lines are the results of the QGSM for the contribution of
the $K^*$ logarithmic Regge trajectory defined by Eq.
(\ref{nonlin}). }
  \label{dstklam2}
\end{figure}

The differential cross section for the reaction $\gamma p \to K^+
\Lambda$ is presented in Figs. \ref{dstklam045}~--~\ref{dstklam2} as
a function of the laboratory photon energy at  fixed values of $t$.
The solid lines are calculated using our model with the following
coupling constants and parameters of the form factor: $g_{\rho K
K^*}=5.8, \ g_{p K^* \Lambda} \simeq 3.5$, $B_1= 5 $~GeV$^{-2}$,
$R^2_1=1.13$~GeV$^{-2}$. The experimental data are from
Refs.\cite{Glander04} (full triangles), \cite{Boyarski69} (full
circles), \cite{Boyarski71}(full stars), \cite{Feller72}(empty
triangles) and \cite{Anderson76} (empty circles). The agreement
between the solid curves and the experimental data clearly supports
the dominant role of the $K^*$ Regge trajectory in the reaction
$\gamma p \to K^+ \Lambda$. The dashed line in Fig.~\ref{dstklam045}
-- calculated at $t=-0.165$ GeV$^2$ -- describes the result for a
$K$ Regge trajectory normalized to the data at $E_{\gamma} = 2$~GeV.
Definitely, it can not describe the energy dependence of the
differential cross section and we may conclude that the $K$ Regge
trajectory is subdominant.

\begin{figure}[t]
  \centerline{\psfig{file=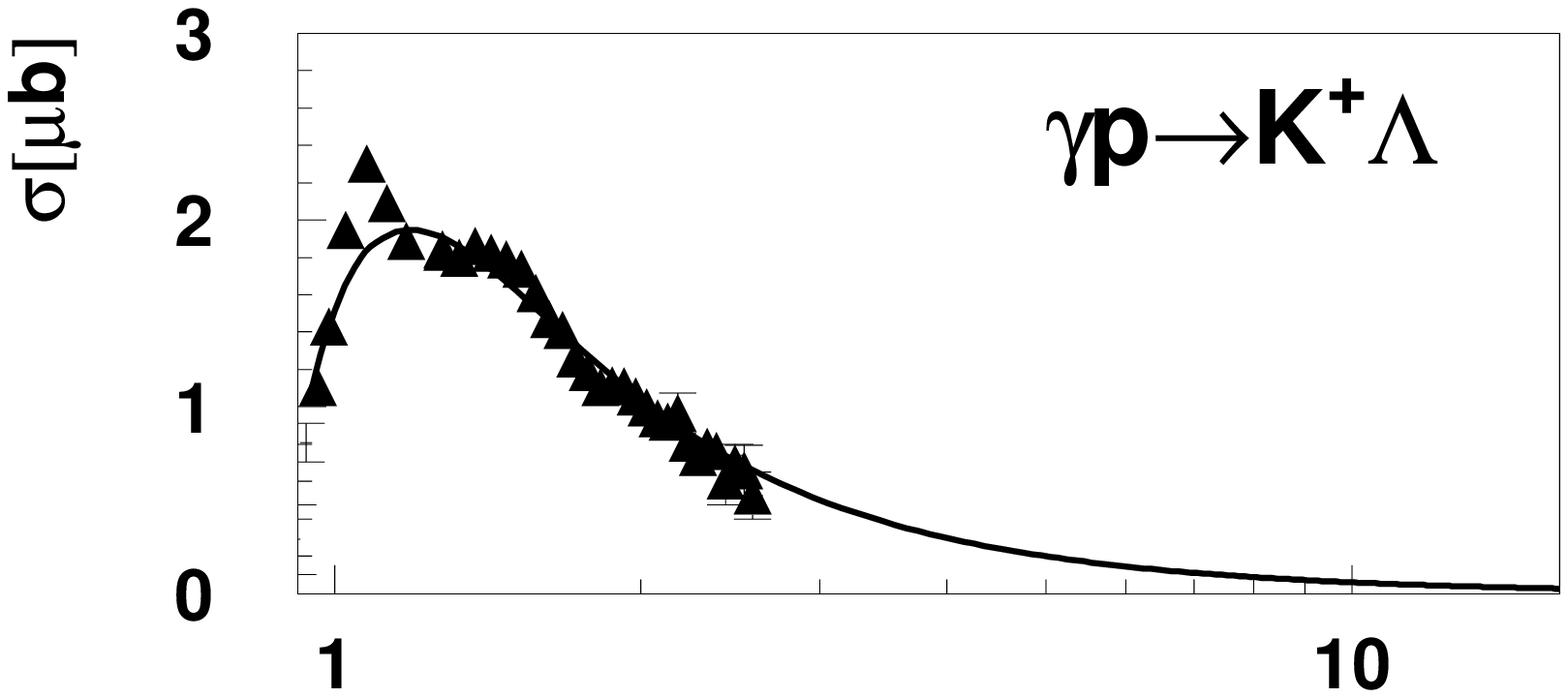,width=7.0cm}}
  \centerline{\psfig{file=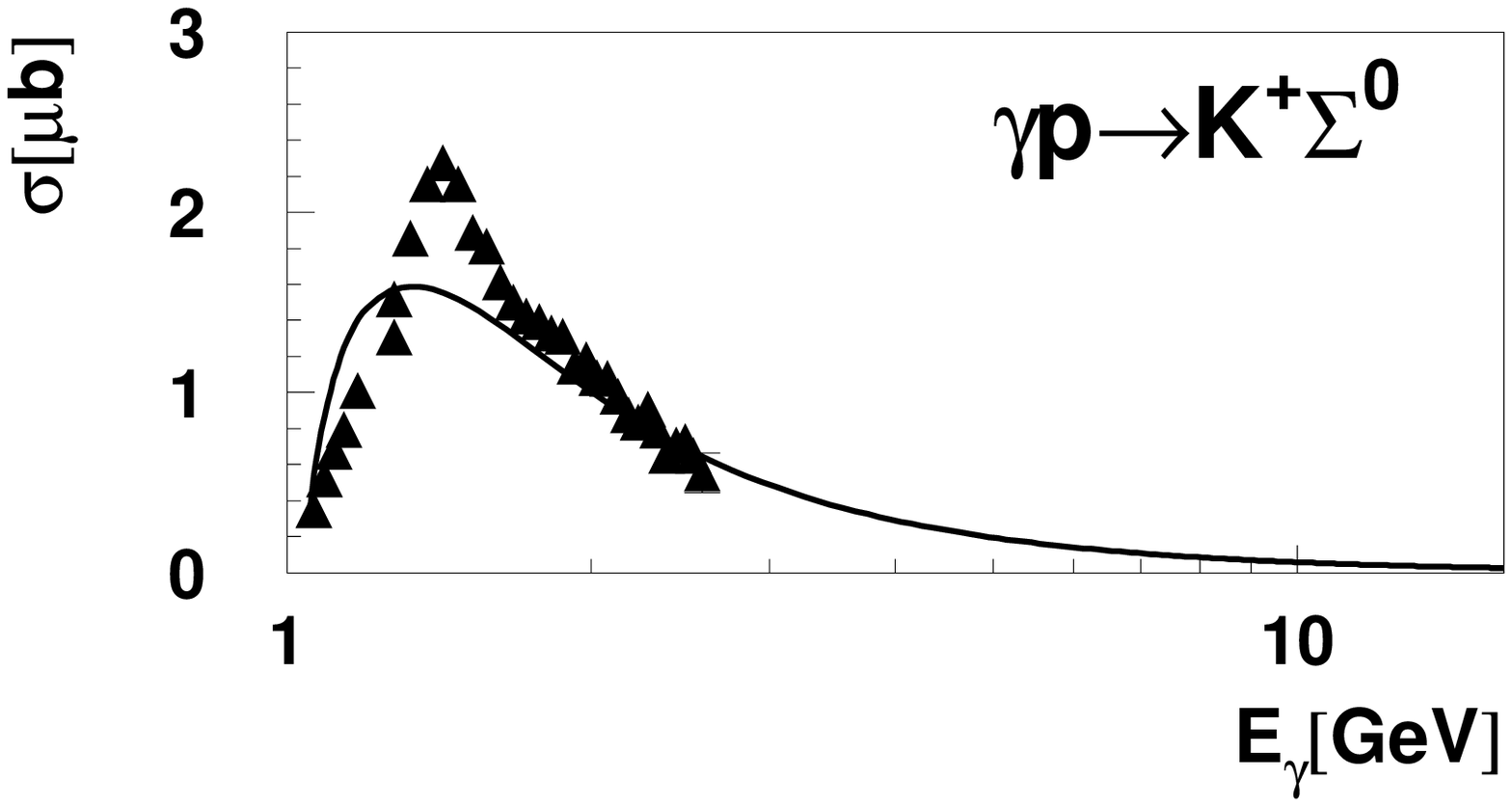,width=7.0cm}}
    \caption{Total cross section of the reactions
    $\gamma p \to  K^+ \Lambda$ and $\gamma p \to  K^+ \Sigma^0$
    as a function of the laboratory photon energy in comparison to the
    experimental data from \cite{Glander04}. The solid line is the result
    of the QGSM for the contribution of the
   $K^*$ logarithmic Regge trajectory defined by Eq. (\ref{nonlin}).}
  \label{slam}
\end{figure}

We have applied our model also to the description of the reaction
$\gamma p \to  K^+ \Sigma^0$ adopting the same coupling constants
and form factor as for the reaction $\gamma p \to  K^+ \Lambda$,
however, modifying the scaling factor to $s_0^{K\Sigma}= (M_{\Sigma}
+ m_{K})^2$. The total cross sections of the reactions $\gamma p \to
K^+ \Lambda$ and $\gamma p \to  K^+ \Sigma^0$ as a function of the
laboratory  photon energy are shown in Fig. \ref{slam} (upper and
lower parts describe the reactions $\gamma p \to  K^+ \Lambda$ and
$\gamma p \to  K^+ \Sigma^0$, respectively) in comparison to the
experimental data  from Ref. \cite{Glander04}. In this context one
has to note that the Regge model gives only the average cross
section for a particular channel and misses resonant amplitudes at
low energy. For example, according to  recent data on the reaction
$\gamma p \to K^+ \Sigma^0$~\cite{McNabb} the $s$-channel resonance
contributions are found to be important for photon beam energies at
least up to 1.5~GeV. Since the $K^+\Lambda$ and $K^+\Sigma^0$
systems show strong resonances in the 1.3 to 2 GeV invariant mass
region the latter cannot be described in the Regge approach.
Nevertheless, the results of our model calculations (presented as
the solid lines) are in a good agreement with the data -- except for
the resonance structures mentioned above -- and support the
'universality' of Reggeon couplings.

\section{Total cross sections for the reactions
$\gamma p \to K^+ \Lambda(1520)$ and $\gamma p \to \bar{K}^0 \Theta
^+$} 
\begin{figure}
 \begin{center}
    \leavevmode
    \psfig{file=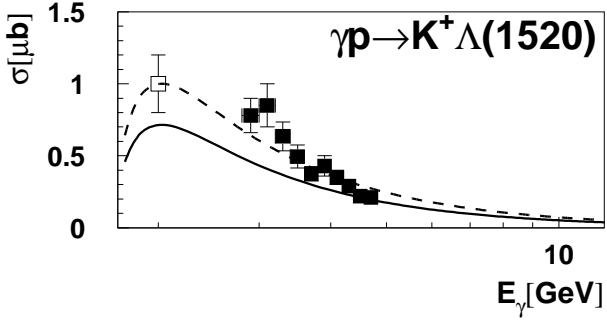,width=8.cm}
    \caption{Total cross section of the reaction $\gamma p \to
    K^+ \Lambda(1520)$
    as a function of the lab. photon energy in comparison to the experimental data
    from Ref. \cite{Barber} (full squares) and Ref. \cite{Barth} (open square).
    The solid line is the result
    of the QGSM for the contribution of the
   $K^*$ logarithmic Regge trajectory defined by Eq. (\ref{nonlin}) (see text);
   the dashed line
   results for a coupling  $g_{pK^* Y}=4.14$.}
 \label{slam1520}
 \end{center}
\end{figure}

In this Section we explore if the universality of the $K^*$
trajectory coupling to
 baryons with constituent $qqs$ quarks also holds in the binary
 reactions
\begin{equation}
\gamma p \to K^+ Y_i
\label{eq:KY}
\end{equation}
with $Y_1= \Lambda(1116), Y_2=\Lambda(1520),...$.
In case of the universality to hold  we have
\begin{equation}
g_{pK^* {\Lambda(1520)}} = g_{pK^* \Lambda} \simeq 3.5,
~F_{\Lambda(1520)}(t)=F_{1}(t) .
\end{equation}
The resulting total cross section of the reaction $\gamma p \to K^+
\Lambda(1520)$ is presented in Fig. \ref{slam1520} in comparison to
the experimental data  from \cite{Barber} (full squares) and
\cite{Barth} (empty square). The solid curve is calculated for the
coupling constant $g_{pK^* Y}=3.5$ which corresponds directly to the
prediction from the universality principle. The dashed curve is
calculated using $g_{pK^* Y}=4.14$ and is in a good agreement with
the data; the deviation between the two curves does not exceed 40\%.
Therefore, the data on the reactions $\gamma p \to K^+ \Lambda$ and
$\gamma p \to K^+ \Lambda(1520)$ support the assumption on the
universality of the $K^*$ trajectory coupling to $q \bar q$ mesons
as well as to baryons with constituent $qqs$ quarks (at least within
a factor of 2). We, accordingly, consider (or define) the
universality principle to hold if a variety of cross sections is
described (predicted) within a factor better than 2.

\begin{figure}
 \begin{center}
    \leavevmode
    \psfig{file=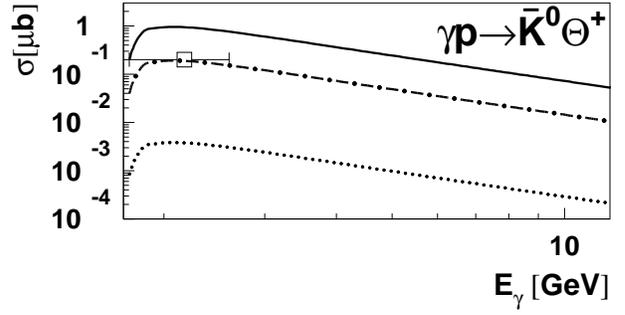,width=8.cm}
    \caption{Total cross section of the reaction
    $\gamma p \to \bar{K}^0 \Theta ^+$.
    The solid line results when assuming the validity of the universality principle
    also for the $\Theta^+$ baryon. The dash-dotted curve
  is calculated for the reduced couplings~(\ref{redu}) assuming the validity of the
  quoted cross section ~(\ref{Theta_cross_section}). The dotted curve is calculated
using the coupling constant $g_{pK^* \Theta}^{\mbox{\tiny
CLAS}}$~(\ref{redu1}) that corresponds to the preliminary upper
limit from the CLAS collaboration (\ref{CLAS_ limit}) on the total
cross section of the reaction $\gamma p \to \bar{K}^0 \Theta ^+$
\cite{DeVita}. In the latter case the 'universality principle' is
found to be violated by almost 3 orders of magnitude. }
 \label{stht}
 \end{center}
\end{figure}

We continue with the $\gamma p \rightarrow \bar{K}^0 \Theta ^+$
reaction and explore if the unversality principle (in the sense
defined above) also holds in this case. The cross section of the
reaction $\gamma p \to \bar{K}^0 \Theta ^+$ was estimated by the
SAPHIR collaboration \cite{Barth} as
\begin{equation}
\sigma_{\gamma p \to \bar{K}^0 {\Theta}^+} \simeq 200 \ {\mbox{nb}}
\label{Theta_cross_section}
\end{equation}
at an average photon energy of $\sim$2 GeV. Let's first adopt 200 nb
as an upper limit for the total cross section of the reaction
$\gamma p \to \bar{K}^0 \Theta ^+$ at 2 GeV. In this case equation
(\ref{Theta_cross_section}) implies already a noticeable violation
of the universality principle for $\Theta^+$ photoproduction since
-- assuming the universality principle to hold for $\Theta^+$ -- we
get
\begin{equation}
 g_{pK^* \Theta} = g_{pK^* {\Lambda(1520)}} \simeq 4.14,~ \ F_2(t)=F_1(t).
\end{equation}
This leads to a total cross section of the reaction $\gamma p \to
\bar{K}^0 \Theta ^+$
 shown by the solid line in Fig.~\ref{stht}. The dash-dotted curve
 in Fig.~\ref{stht} is calculated assuming
\begin{equation}
g_{pK^* \Theta}^{\mbox{\tiny SAPHIR}} \simeq  0.4 \  g_{pK^*{\Lambda(1520)} }
\simeq 1.8 \ , \ F_2(t)=F_1(t)
\label{redu}
\end{equation}
in order to match the quoted cross section in
(\ref{Theta_cross_section}).  However, the new preliminary results
from the CLAS collaboration \cite{DeVita} do not support the
estimate (\ref{Theta_cross_section})and indicate that the upper
limit on the total cross section of the reaction $\gamma p \to
\bar{K}^0 \Theta ^+$ should be much lower:
\begin{equation}
 \sigma_{\gamma p \to \bar{K}^0 \Theta ^+} \leq 1\div 4 \ \mbox{nb}\ .
\label{CLAS_ limit}
\end{equation}
When taking 4~$nb$ as an upper limit we find
\begin{equation}
g_{pK^* \Theta}^{\mbox{\tiny CLAS}} \simeq  0.06 \  g_{pK^*{\Lambda(1520)} }
\simeq 0.25 \ , \ F_2(t)=F_1(t) \ .
\label{redu1}
\end{equation}
Therefore, using the preliminary result from the CLAS collaboration
we find a very strong suppression of the $\gamma p \to \bar{K}^0
\Theta ^+$ cross section relative to the prediction from the
universality principle for photoproduction of the lowest $qqs$
baryons. If the pentaquark exists, we may interpret this finding as
a clear indication of a substantially different quark structure of
the $\Theta^+$.

\section{Pion induced reactions:
$\pi^- p \to K^0 \Lambda$ and $\pi^- p \to K^- \Theta ^+$ }
\begin{figure}
 \begin{center}
    \leavevmode
    \psfig{file=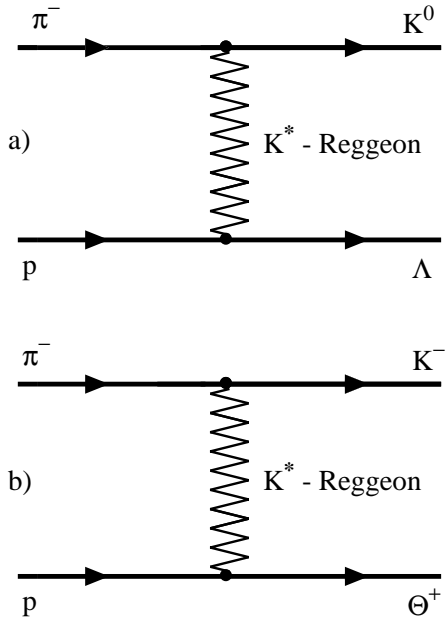,width=6.cm}
    \caption{$K^*$ Reggeon exchanges corresponding to the
quark diagrams of  Fig. \ref{fig:qdiagr}.}
\label{fig:pipklkt}
  \end{center}
\end{figure}

\begin{figure}[t]
  \centerline{\psfig{file=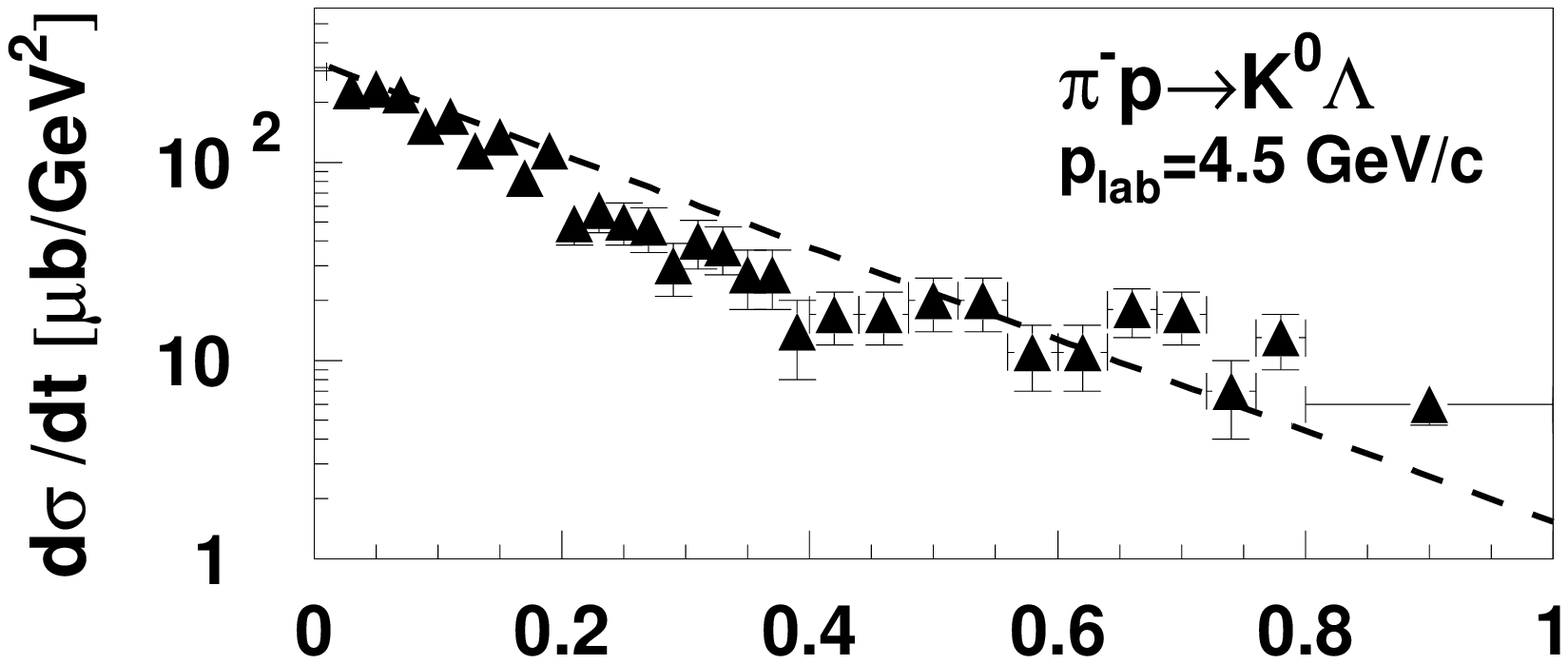,width=7.0cm}}
  \centerline{\psfig{file=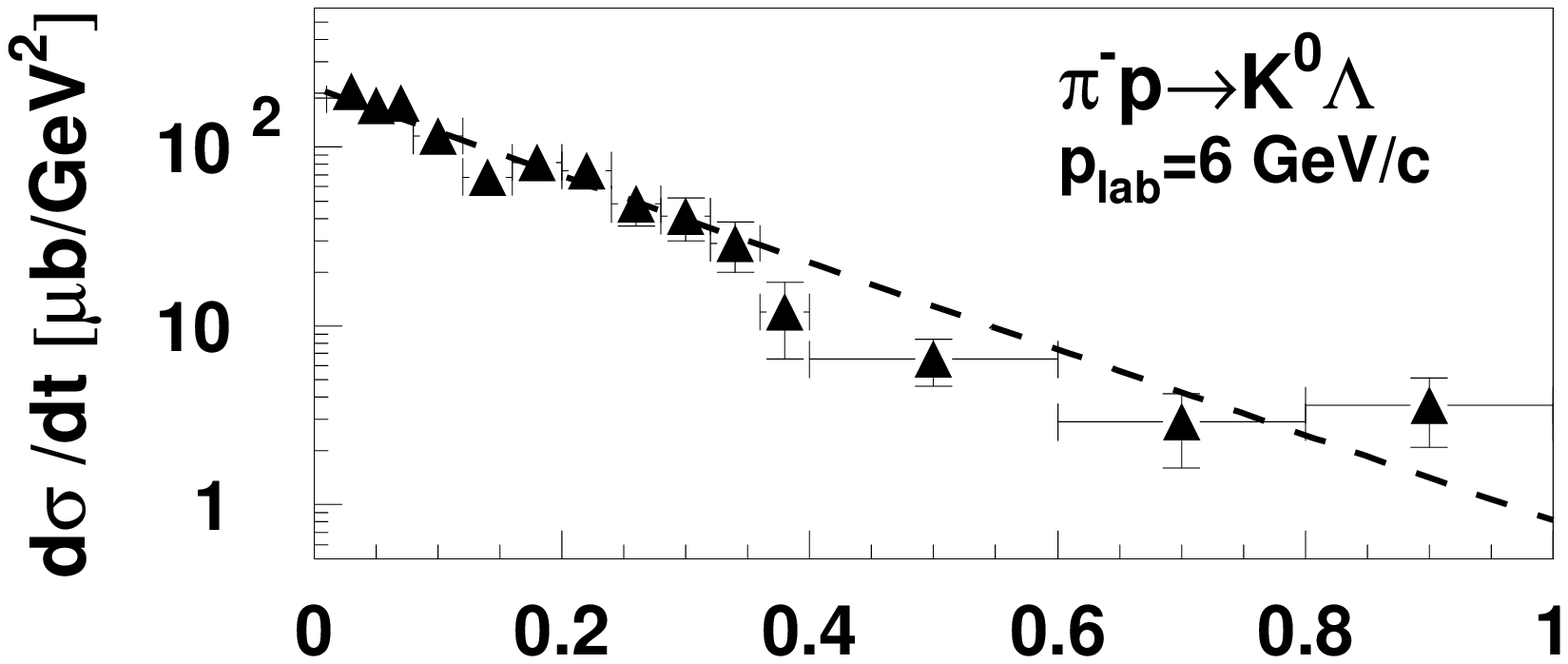,width=7.0cm}}
  \centerline{\psfig{file=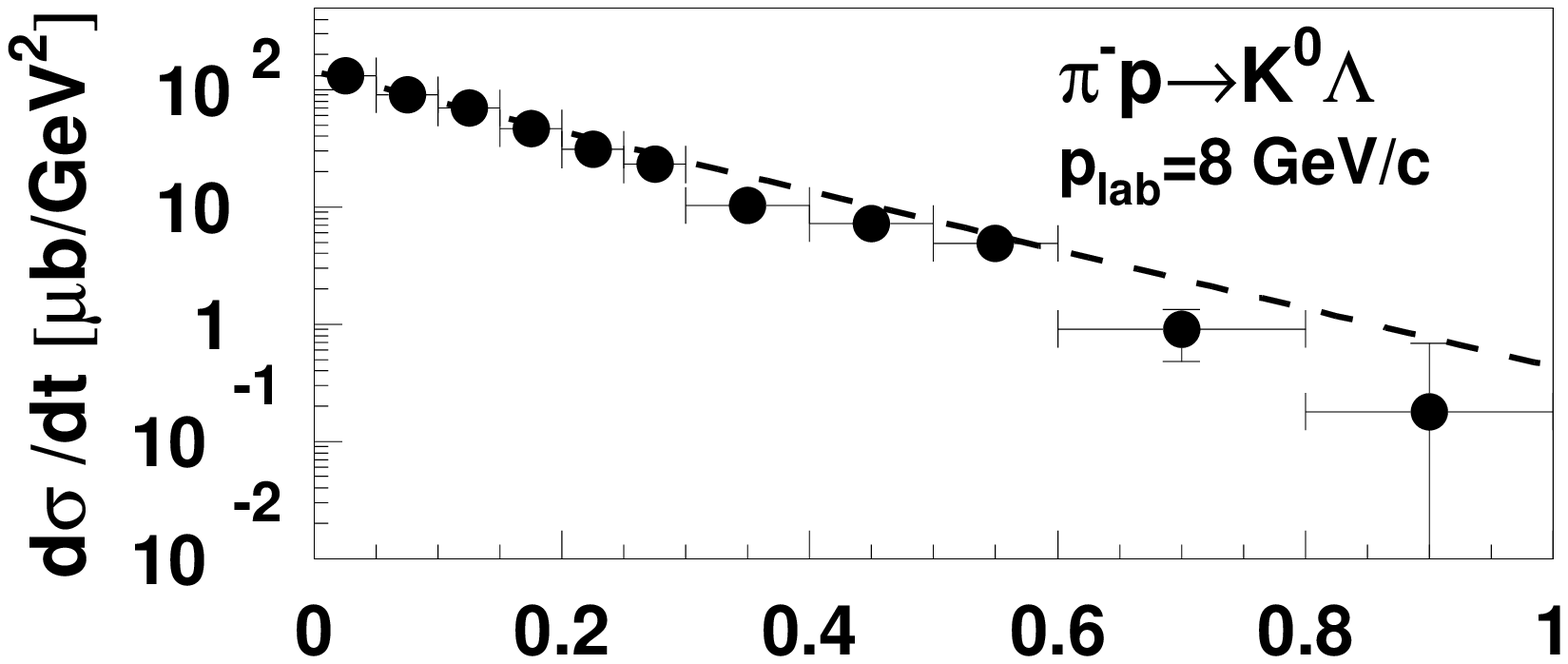,width=7.0cm}}
  \centerline{\psfig{file=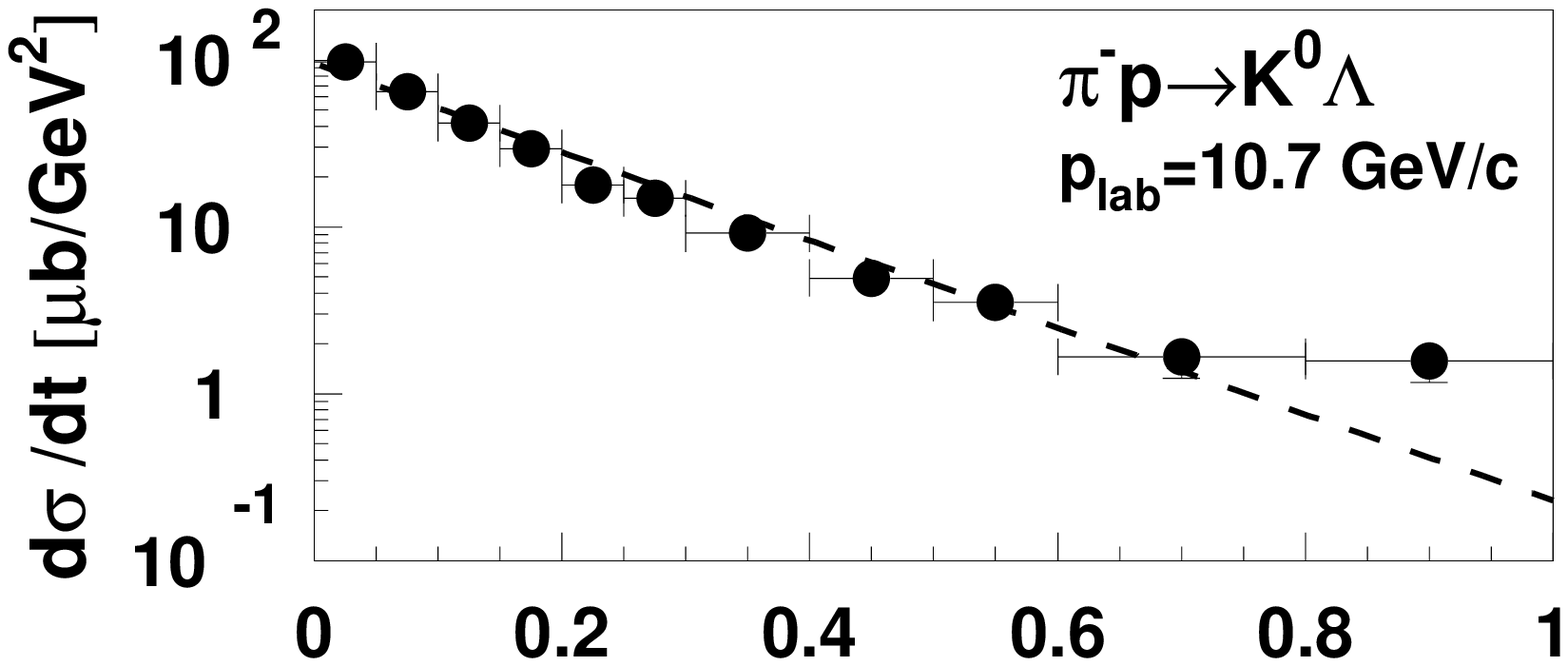,width=7.0cm}}
  \centerline{\psfig{file=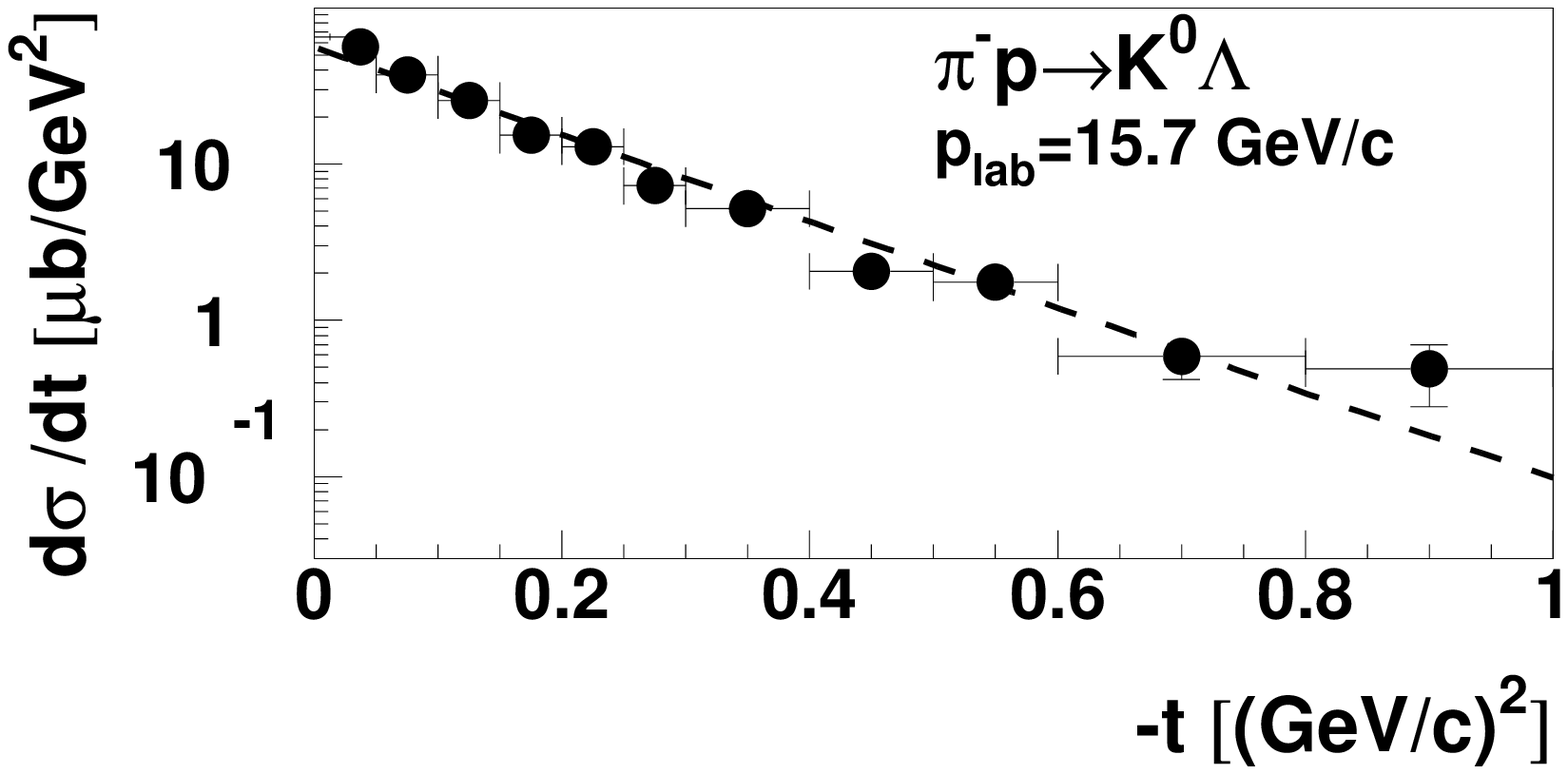,width=7.0cm}}
    \caption{The differential cross section
of the  $\pi^- p \to K^0 \Lambda$ reaction as a function
of $t$  at different laboratory momenta p$_{\mathrm{lab}}$ in comparison to the
experimental data from \cite{Foley73} (full circles)
and~\cite{Crennel} (full triangles).
The results of our model are shown by the dashed lines.}
  \label{dstpipkl}
\end{figure}

We continue with $\pi^-$ induced reactions and assume that the  amplitudes
of the reactions $\pi^- p \to K^0 \Lambda$
and $\pi^- p \to {K^-} \Theta ^+$  are also dominated by the contribution
of the $K^*$ Regge trajectory (see Fig.~\ref{fig:pipklkt} a) and
b)) such that the cross sections are fully determined. We directly step on with the results
for the differential cross section
of the  $\pi^- p \to K^0 \Lambda$ reaction as a function
of $t$ at $p_{\mathrm{lab}}$=4.5, 6, 8, 10.7 and 15.7~GeV/c shown in
 Fig. \ref{dstpipkl} in comparison to the experimental data
from~\cite{Foley73,Crennel}.
The results of our model are displayed by the dashed lines calculated for the following
parameters:
\begin{equation}
g_{\pi K^* K} =5.8, ~g_{pK^* \Lambda} \simeq 4.5,
~B=0,~R_1^2=2.13~ \mathrm{GeV^{-2}} \label{pipkl_g}.
\end{equation}
We see that the coupling constants $g_{\pi K^* K}$ and $g_{pK^* \Lambda}$
-- in the case of the reaction $\pi^- p \to K^0 \Lambda$ -- are also in agreement
with the universality assumption for the coupling constants.
However, to describe the $t$-dependence of
the differential cross section we have to employ different parameters
for the form factor $F(t)$ as compared to the reaction
$\gamma p \to {K^+} \Lambda$. This can be explained, in particular,
by the different relative contributions of the baryon spin-flip terms in
the reactions  $\gamma p \to K^+ \Lambda$ and
$\pi^- p \to K^0 \Lambda$.

\begin{figure}
 \begin{center}
    \leavevmode
    \psfig{file=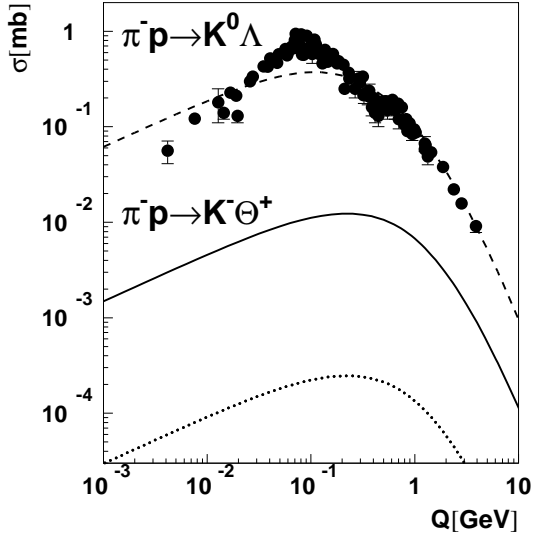,width=7.cm}
    \caption{Total cross section of the reaction
    $\pi^- p \to K^0 \Lambda$ as a function of the c.m. excess
    energy $Q= \sqrt{s} - m_K - M_{\Lambda}$ (dashed line) in comparison to the data
  from Ref.~\cite{Landoldt}. The solid and dotted line show the expected total
  cross section of the reaction
    $\pi^- p \to K^- \Theta^+$ as a function of the c.m. excess
    energy $Q= \sqrt{s} - m_K - M_{\Theta^+ (1540)}$
 calculated with the coupling constants
$g_{pK^* \Theta}=g_{pK^* \Theta}^{\mbox{\tiny SAPHIR}}$ (19) and
$g_{pK^* \Theta}^{\mbox{\tiny CLAS}}$ (21). }
 \label{pipkltot}
 \end{center}
\end{figure}

The total cross section of the reaction $\pi^- p \to K^0 \Lambda$  is presented in
Fig.~\ref{pipkltot} where the dashed curve is the result of our calculations.
As expected for a Regge model  -- and discussed above --
we see some deviation of the QGSM from the data~\cite{Landoldt} in the resonance region.
However, at higher energies, where the explicit resonance
structure disappears, the theoretical calculations are in a good agreement
with the data.

Now using the coupling constants $g_{\pi K^* K} =5.8$, $g_{pK^*
\Theta}=g_{pK^* \Theta}^{\mbox{\tiny SAPHIR}}$ and $g_{pK^*
\Theta}^{\mbox{\tiny CLAS}}$ and assuming $R_1^2$ and $B$ to be the
same as for the reaction  $\pi^- p \to K^0 \Lambda$  we can
calculate the cross section for the reaction $\pi^- p \to K^-
\Theta^+$ . The results are shown in Fig.~\ref{pipkltot} by the
solid and dotted line, respectively; these cross sections reach
about 10 (0.2)~$\mu$b in their maximum.

\section{$\Theta^+$ production in exclusive and
inclusive $NN$  collisions} 
\begin{figure}
 \begin{center}
    \leavevmode
    \psfig{file=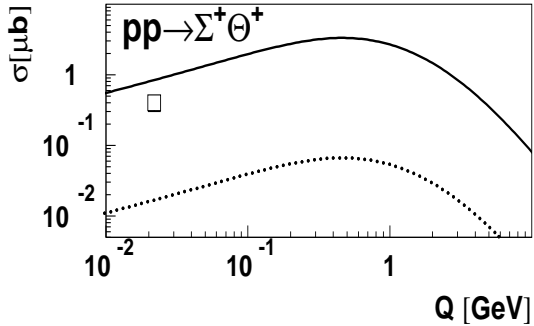,width=7.cm,height=4.3cm}
    \caption{Total cross section for $\Theta^+$
production as a function of the excess energy $Q$ in the reaction
$pp \to \Theta^+ \Sigma^+$. The solid (dotted) line is the result of
a calculation with the coupling constants $g_{p K^* \Sigma}=g_{p K^*
\Lambda}=4.5$, $g_{pK^* \Theta}=g_{pK^* \Theta}^{\mbox{\tiny
SAPHIR}}$ ($g_{pK^* \Theta}^{\mbox{\tiny CLAS}}$). The data point
(open square) is from the COSY-TOF collaboration~\cite{Abdel-Bary}.
}
 \label{ppthetsig}
 \end{center}
\end{figure}

\begin{figure}[t]
  \centerline{\psfig{file=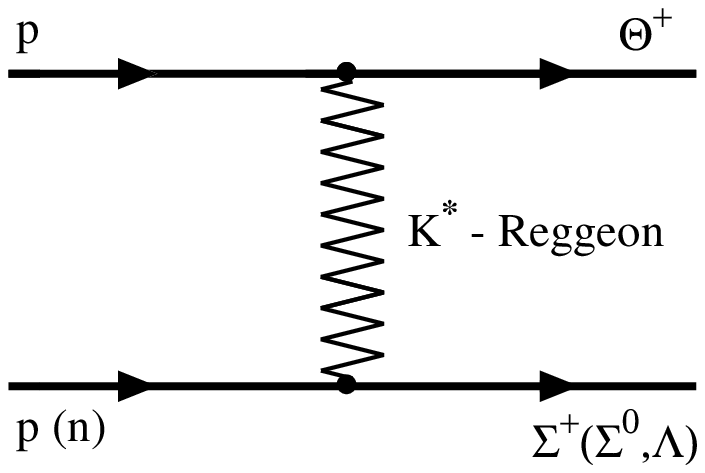,width=5.0cm}}
  \centerline{\psfig{file=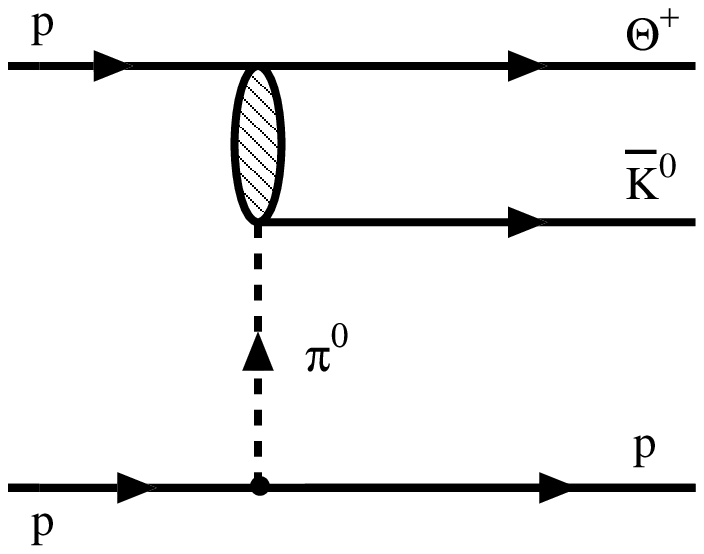,width=5.0cm}}
  \centerline{\psfig{file=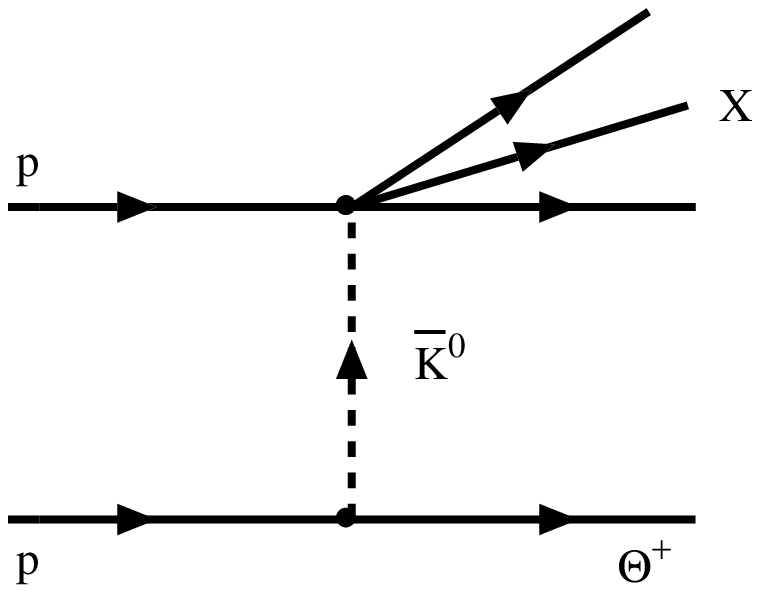,width=5.0cm}}
     \caption{The diagrams describing the $\Theta^+$
production in the reactions: $pp \to \Theta^+ \Sigma^+$
or $pn \to \Theta^+ \Sigma^0 (\Lambda)$ (upper diagram),
$pp \to \Theta^+ \bar{K}^0 p$ with pion exchange (middle diagram)
and $pp \to \Theta^+ X$ with kaon exchange (lower diagram).}
  \label{fig:ppthetax}
\end{figure}

Further constraints on the universality
of amplitudes and cross sections for $\Theta^+$ production are
provided by $NN$ collisions. Here the amplitude of the reaction $pp
\to \Theta^+ \Sigma^+ $ (cf. upper diagram in
Fig.~\ref{fig:ppthetax}) has been calculated using the coupling
constant $g_{pK^* \Theta}=g_{pK^* \Theta}^{\mbox{\tiny SAPHIR}}$
($g_{pK^* \Theta}^{\mbox{\tiny CLAS}}$). Of course, taking into
account the results of Section~2 we can safely assume that $g_{pK^*
\Sigma}\simeq g_{pK^* \Lambda}$. At the same time the coupling
constant $g_{pK^* \Lambda}$  may vary from 3.5 (if we define it via
the reaction  $\gamma p \to K^+ \Lambda$) to 4.5 (if we define it
via the $\pi^- p \to K^0 \Lambda$ reaction ). In this section we use
$g_{pK^* \Lambda}$  = 4.5. The predictions for the total cross
section of the $pp \to \Theta^+ \Sigma^+ $ reaction are shown by the
solid (dotted) line in Fig.~\ref{ppthetsig}
 using  $g_{pK^* \Theta}=g_{pK^* \Theta}^{\mbox{\tiny SAPHIR}}$
($g_{pK^* \Theta}^{\mbox{\tiny CLAS}}$) . The cross section
described by the dotted line is $\sim$ 20--30 times smaller than the
experimental value $\sigma=0.4\pm 0.1(\mbox{stat}) \pm 0.1
(\mbox{syst})~\mu$b~\cite{Abdel-Bary} measured at a beam momentum of
$2.95$~GeV/c (open square) and clearly signals an incompatibility of
the different measurements.

We note, that the first analysis of the reaction $pp \to \Theta^+
\Sigma^+ $ was performed by Polyakov et al. in Ref. \cite{Polyakov}
even before the $\Theta^+$ baryon was 'discovered' from the
experimental side. Their estimation of the cross section -- within
the kaon-exchange approximation -- was  about 2 $\mu$b at the
initial momentum $\sim$3 GeV/c. Approximately the same value of the
total cross section of the reaction $pp \to \Theta^+ \Sigma^+ $ was
found by Liu and Ko in Ref. \cite{LiuKo} later on. Our calculated
cross section is about 0.8~$\mu$b on the basis of the coupling (19)
and 16 $nb$ for the coupling (21) at the excess energy $Q=22$~MeV
($p_{\mathrm{lab}}=2.95$~GeV/c). The maxima of cross sections are
$3.3 \ \mu$b and 66$~nb$, respectively, at the excess energy
$Q=460$~MeV ($p_{\mathrm{lab}}=4.4$~GeV/c). These cross sections are
smaller by factors  of 2 and 90, respectively, compared to the early
estimates.

\begin{figure}
 \begin{center}
    \leavevmode
    \psfig{file=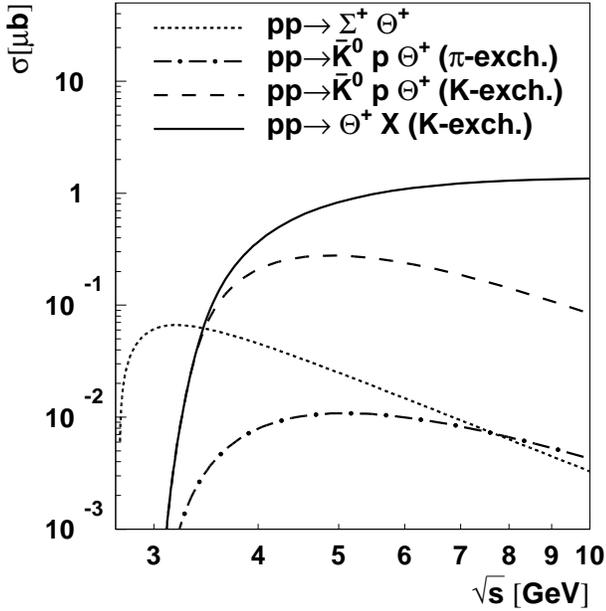,width=8.cm}
    \caption{Total cross sections for $\Theta^+$
production as a function of the c.m. energy in the reactions:
$pp \to \Theta^+ \Sigma^+$ (dotted curve) and
$pp \to \Theta^+ \bar{K}^0 p$ with pion exchange (dash-dotted line)
calculated with the coupling constant
$g_{p K^* \Theta}=g_{pK^* \Theta}^{\mbox{\tiny CLAS}}$ from
Eq.~(\ref{redu1}),
$pp \to \Theta^+ \bar{K}^0 p$ with kaon exchange (dashed line),
and $pp \to \Theta^+ X$ with kaon exchange (solid line).}
 \label{ppthetax}
 \end{center}
\end{figure}

We continue with alternative production mechanisms for $\Theta^+$ production
in $pp$ collisions as described by
the middle and lower diagrams in Fig.~\ref{fig:ppthetax}.
In the following we use the method of Yao~\cite{Yao} to calculate the
cross sections of the reactions: \\
1) $pp\to p \bar {K}^0 \Theta^+$ with the pion exchange
(middle diagram in Fig.~\ref{fig:ppthetax})\\
2) $pp\to p \bar {K}^0 \Theta^+$ with the kaon exchange (lower
diagram in Fig.~\ref{fig:ppthetax}
with $X=\bar {K}^0 p$), \\
3) $pp\to \Theta^+ X$ with the kaon exchange
(lower diagram in Fig.~\ref{fig:ppthetax}).\\
In the case of pion exchange the expression for the
total cross section can be written in the
form:
\begin{eqnarray}
&&  \sigma (pp \to p \bar{K}^0 \Theta^+)= \nonumber \\
&&\frac{G^2_{\pi NN}}{8 {\pi}^2 p_{1} s}
\int_{W_{\mathrm {min}}}^{W_{\mathrm {max}}} k \, W^2 \
\sigma ({\pi}^0 p \to \bar{K}^0 {\Theta}^+, W)\ dW \nonumber \\
&&\times \int_{t_{\mathrm{min}}(W)}^{t_{\mathrm{max}}(W)}
F_{\pi}^4(t)\ \frac{1}{(t-m_{\pi}^2)^2} \ t\, dt \ ,
\label{pppiex}
\end{eqnarray}
where $W$ is the invariant mass of the $\bar{K}^0 {\Theta}^+$
system, $k$ is defined as
$$
k=\left((W^2-(m_p-m_{\pi})^2)(W^2-(m_p+m_{\pi})^2)\right)^{1/2}/2W \ ,
$$
$p_1$ is the initial proton momentum in the c.m. system, $t=(p_2-p_4)^2$,
and $G_{\pi NN}=13.45$.
Assuming that $J^P(\Theta^+)=\frac12^+$ we have the following
expression for the kaon exchange contribution
\begin{eqnarray}
&&  \sigma (pp \to {\Theta}^+ X)= \nonumber \\
&&\frac{G^2_{\Theta KN}}{8 {\pi}^2 p_{1} s}
\int_{W_{\mathrm {min}}}^{W_{\mathrm {max}}} k \, W^2 \
\sigma (\bar{K}^0 p \to X, W)\ dW \nonumber \\
&&\times \int_{t_{\mathrm{min}}(W)}^{t_{\mathrm{max}}(W)}
F_{K}^4(t)\ \frac{1}{(t-m_{K}^2)^2} \ \left(t-\Delta_{M43}^2
\right)\, dt \ .
\label{ppkex}
\end{eqnarray}
Here $W$ is the invariant mass of the system $X$ and
$\Delta_{M43}^2=(m_{\Theta}-m_p)^2$.
The form-factors for the virtual pion and kaon exchange have been
 chosen of the monopole type
\begin{equation}
F_j(t)=\frac{\Lambda_j^2-m_j^2}{\Lambda_j^2-t} \,
\end{equation}
with $\Lambda_{\pi}=1.3$~GeV and $\Lambda_K=1$~GeV. These parameters
have been used in Ref.~\cite{Gasparyan} to describe the total cross
section of the reaction $pp \to K^+ \Lambda p$. Obviously, the
contribution of the kaon exchange to the reaction $pp \to \Theta^+
X$ depends on the coupling constant $G_{\Theta KN}$, which can only
be estimated (or fixed by upper limits). If the ${\Theta}^+$ decay
width is less than $1$~MeV~\cite{PDG} we have $G_{\Theta KN} \leq
1.4$. The solid and dashed lines in Fig.~\ref{ppthetax} present our
results for the inclusive $pp\to \Theta^+ X$ and exclusive $pp\to
\Theta^+ \bar{K}^0 p$ reactions as calculated within the
kaon-exchange model for $G_{\Theta KN} = 1.4$ (as an upper limit).
The inclusive cross section turns out to be about $1.5~\mu$b at high
energies while the exclusive $pp\to \Theta^+ \bar{K}^0 p$ cross
section is at least one order of magnitude smaller.

The reaction $pp\to p \bar {K}^0  \Theta^+$ may proceed also via
$\pi$ exchange. The corresponding contribution to the total cross
section calculated with the coupling constant $g_{p K^*
\Theta}=g_{pK^* \Theta}^{\mbox{\tiny CLAS}}$ from Eq.~(\ref{redu1})
is presented by the dash-dotted line in Fig.~\ref{ppthetax}. It
reaches  about $10$~nb in its maximum. Therefore, we can conclude
that the $\Theta^+$ production cross section in $pp$-collisions is
dominated by the $K$ exchange and should be about a few $\mu$b. Note
that using the hadronic Lagrangian model Liu and Ko \cite{LiuKo}
found a considerably larger cross section for $\Theta^+$ production,
i.e. about 50 $\mu$b in pion-nucleon reactions and $\sim$20 $\mu$b
in proton-proton reactions. From our point of view the latter
results are essentially due to a large coupling constant $G_{\Theta
KN} \simeq 4.4$ in \cite{LiuKo}, which corresponds to
$\Gamma_{\Theta} \approx 20$ MeV. Such large couplings, however,
should be excluded according to the more recent analysis in
Ref.~\cite{Sibirtsev}.

\section{Conclusions}
In this study we have analyzed $\Lambda$, $\Sigma^0$,
$\Lambda(1520)$ and $\Theta^+$ production in binary reactions
induced by photon, pion and proton beams in the framework of the
Quark-Gluon Strings Model combined with Regge phenomenology.
Starting with the existing experimental data on the $\gamma p \to
K^+ \Lambda$ reaction we have demonstrated that the differential and
total cross sections at photon energies $1 - 16$~GeV and $-t <
2$~GeV${}^2$ can be described very well by the model with a dominant
contribution of the $K^*$ Regge trajectory. We stress  that the
rather good description of the large $t$ region was possible only
due to  the logarithmic form of the $K^*$ Regge trajectory
(\ref{nonlin}). It has been demonstrated, furthermore, that the data
on the reactions $\gamma p \to K^+ \Lambda$, $\gamma p \to K^+
\Sigma^0$ and $\gamma p \to K^+ \Lambda(1520)$  -- at least within a
factor of 2 --  support the assumption on the universality of the
$K^*$ trajectory coupling to $q \bar q$ mesons as well as to baryons
with $qqs$ constituent quarks. This implies that -- using the same
parameters as for $\gamma p \to K^+ \Lambda$ -- we are able to
reproduce the total $\gamma p \to K^+ \Sigma^0$ and $\gamma p \to
K^+ \Lambda(1520)$ cross sections (within an accuracy of 40\%). On
the other hand, as a consequence of the SAPHIR data \cite{Barth} and
the preliminary data from CLAS \cite{DeVita}, there is an essential
suppression of the $\gamma p \to \bar{K}^0 \Theta ^+$ cross section
relative to the prediction within the universality principle that
was shown to hold (with reasonable accuracy) for the photoproduction
of the lowest $qqs$ baryons. We conclude that this suppression
indicates a substantially different quark structure and wave
function of the $\Theta^+$ (in case of its final experimental
confirmation).

Moreover, we have suggested that the  amplitudes of the reactions $\pi^- p \to K^0 \Lambda$
and $\pi^- p \to {K^-} \Theta ^+$  are also dominated by the contribution
of the $K^*$ Regge trajectory (cf. Fig.\ref{fig:pipklkt} a) and b)).
Indeed, the differential and total cross sections
of the  $\pi^- p \to K^0 \Lambda$ reaction are found
to be in a reasonable agreement with the universality principle.
Using parameters defined by the  analysis of the reactions
$\gamma p \to K^+ \Lambda$, $\gamma p \to K^+ \Sigma^0$, $\gamma p \to K^+ \Lambda(1520)$,
$\gamma p \to \bar{K}^0 \Theta^+$ and
 $\pi^- p \to K^0 \Lambda$ we have calculated the cross section
for the reaction  $\pi^- p \to {K^-} \Theta ^+$ . We predicted a maximum
cross section of about 200~nb (cf. Fig. 10) for
the coupling constant $g_{pK^* \Theta}=g_{pK^* \Theta}^{\mbox{\tiny CLAS}}$
extracted from the new preliminary CLAS result
on the reaction $\gamma p \to \bar{K}^0 \Theta^+$~\cite{DeVita}.

We extended our model additionally to the analysis of the binary
reaction $pp \to \Sigma^+ \Theta^+$. We found that the cross section
of this reaction -- measured by the COSY-TOF collaboration
\cite{Abdel-Bary}-- is 20--30 times larger than the value predicted
by the model with the coupling constant $g_{pK^* \Theta}=g_{pK^*
\Theta}^{\mbox{\tiny CLAS}}$. Furthermore, we have investigated the
exclusive and inclusive $\Theta^+$ production in the reactions $pp
\to p \bar{K}^0 \Theta^+$ and $pp \to \Theta^+ X$ and found that the
inclusive $\Theta^+$ production in pp collisions at high energy
should be on the level of 1 $\mu$b.

The systematic and comparative Regge analysis -- provided by our
study -- will also allow in future to relate different data sets
from $\gamma$, $\pi$ and proton induced reactions on $s\bar{s}$ pair
production and finally should yield a transparent picture of the
dynamics as well as the properties of (possible) exotic states.

\section*{Acknowledgments}
The authors are grateful to A.B. Kaidalov and E.~De~Sanctis for
useful discussions. This work was partially supported by DFG
(Germany) and INFN (Italy). One of us (V. G.) acknowledges the
financial support from For\-schungszentrum J\"ulich (FFE grant
41520739 (COSY - 071)).

\end{document}